%% file: main.tex
\makeatletter
\def\input@path{{Styles/}}
\makeatother
\documentclass[sigconf, nonacm]{acmart}
\usepackage[a-2b, mathxmp]{pdfx}
\makeatletter
\newif\if@restonecol
\makeatother

\usepackage[ruled,vlined]{algorithm2e}
\usepackage{amsmath}

\usepackage{enumitem}
\usepackage{multicol}
\usepackage{multirow}
\usepackage{float}
\usepackage{subfig}
\usepackage{balance}
\usepackage{bm}
\usepackage{cleveref}
\usepackage[table]{xcolor}
\usepackage[normalem]{ulem}
\usepackage{colortbl}
\usepackage{booktabs}     
\usepackage{makecell}
\usepackage{booktabs}
\usepackage{tabularx}
\usepackage{CJKutf8}
\usepackage{array}
\usepackage{pifont}
\usepackage{bbding}
\usepackage{graphicx}
\usepackage{tcolorbox}
\usepackage{listings}

\theoremstyle{Example}

\input{macros}

\AtBeginDocument{%
  }

\newcommand\vldbdoi{XX.XX/XXX.XX}
\newcommand\vldbpages{XXX-XXX}
\newcommand\vldbvolume{19}
\newcommand\vldbissue{12}
\newcommand\vldbyear{2026}
\newcommand\vldbauthors{\authors}
\newcommand\vldbtitle{\shorttitle} 
\newcommand\vldbavailabilityurl{}
\newcommand\vldbpagestyle{empty}
\renewcommand\vldbavailabilityurl{https://github.com/petrizhang/relify}

\begin{document}


\input{metadata}
\maketitle
\vspace{-1.5em}

\pagestyle{\vldbpagestyle}
\begingroup\small\noindent\raggedright\textbf{PVLDB Reference Format:}\\
\vldbauthors. \vldbtitle. PVLDB, \vldbvolume(\vldbissue): \vldbpages, \vldbyear.\\
\href{https://doi.org/\vldbdoi}{doi:\vldbdoi}
\endgroup

\vspace{-.5em}

\ifdefempty{\vldbavailabilityurl}{}{
\vspace{.3cm}
\begingroup\small\noindent\raggedright\textbf{PVLDB Artifact Availability:}\\
The source code, data, and/or other artifacts have been made available at \url{\vldbavailabilityurl}.
\endgroup
}

\begingroup
\renewcommand\thefootnote{}\footnote{\noindent
This work is licensed under the Creative Commons BY-NC-ND 4.0 International License. Visit \url{https://creativecommons.org/licenses/by-nc-nd/4.0/} to view a copy of this license. For any use beyond those covered by this license, obtain permission by emailing \href{mailto:info@vldb.org}{info@vldb.org}. Copyright is held by the owner/author(s). Publication rights licensed to the VLDB Endowment. \\
\raggedright Proceedings of the VLDB Endowment, Vol. \vldbvolume, No. \vldbissue\ %
ISSN 2150-8097. \\
\href{https://doi.org/\vldbdoi}{doi:\vldbdoi} \\
}\addtocounter{footnote}{-1}\endgroup

\renewcommand{\shortauthors}{Wu et al.}

\input{Sections/1-Introduction}
\input{Sections/2-Preliminaries}

\input{Sections/3-Overview}
\input{Sections/4-Storage}
\input{Sections/5-Query}
\input{Sections/6-Plan}
\input{Sections/7-Experiments}
\input{Sections/8-Conclusion}

\vspace{-.5em}
\begin{acks}
\begin{sloppypar}
Xuanhe Zhou is the corresponding author. Xufei Wu conducted this work during his Tencent internship, where Pengcheng Zhang was his mentor.
This work was supported in part by National Key R\&D Program of China (No. 2023YFB4502400), 
China NSF grant (No. 62441236, 62372296, 62432007, U25A6024, U25A20437, 62525202, 62232009, 62502304), 
Fundamental and Interdisciplinary Disciplines Breakthrough Plan of the Ministry of Education of China (No. JYB2025XDXM103), 
Tencent Rhino Bird Key Research Project, Shenzhen Project (CJGJZD20230724093403007), Zhongguancun Lab, and Beijing National Research Center for Information Science and Technology (BNRist), Tencent, Shanghai Jiao Tong University AI for Engineering Initiative.

\end{sloppypar}
\end{acks}

\clearpage

\bibliographystyle{ACM-Reference-Format}
\balance
\bibliography{reference}

\end{document}

%% file: macros.tex
\newcommand{\ourSys}{\texttt{TEngineDB-V}\xspace}
\newcommand{\ourPQ}{\texttt{DPPQ}\xspace}
\newcommand{\ourIVFPQ}{\texttt{IVFDPPQ}\xspace}

\newcommand{\hi}[1]{\vspace{.25em} \noindent {\bf #1}\xspace}

\newcommand{\sym}[2][1.25]{\raisebox{0.15ex}{\scalebox{#1}{#2}}}
\newcommand{\cmark}{\sym{\textcolor{green!60!black}{\ding{51}}}} 
\newcommand{\xmark}{\sym{\textcolor{red!70!black}{\ding{55}}}} 

\definecolor{deepPink}{RGB}{204,0,102}

%% file: metadata.tex
\title{\ourSys: An OLAP-Native Vector Search System for Large-$k$ Workloads at Tencent}

\author{Xufei Wu}
\authornote{Equal contribution.}
\affiliation{
  \institution{Shanghai Jiao Tong Univ.}
  \country{}
}
\email{wuxufei0807@gmail.com}

\author{Pengcheng Zhang}
\authornotemark[1]
\affiliation{
  \institution{Tencent Inc.}
  \country{}
}
\email{petrizhang@tencent.com}

\author{Yitong Song}
\affiliation{
  \institution{Hong Kong Baptist Univ.}
  \country{}
}
\email{ytsong@comp.hkbu.edu.hk}

\author{Xiaobo Zhang}
\affiliation{
  \institution{Shanghai Jiao Tong Univ.}
  \country{}
}
\email{strange_uncle@sjtu.edu.cn}

\author{Anqi Liang}
\affiliation{
  \institution{HKUST(GZ)}
  \country{}
}
\email{anqiliang9657@gmail.com}

\author{Kai Wang}
\affiliation{
  \institution{Shanghai Jiao Tong Univ.}
  \country{}
}
\email{w.kai@sjtu.edu.cn}

\author{Jijun Du}
\affiliation{
  \institution{Tencent Inc.}
  \country{}
}
\email{jijundu@tencent.com}

\author{Yidi Xiong}
\affiliation{
  \institution{Tencent Inc.}
  \country{}
}
\email{eddyxiong@tencent.com}

\author{Guangxu Cheng}
\affiliation{
  \institution{Tencent Inc.}
  \country{}
}
\email{andrewcheng@tencent.com}

\author{Zhe Chen}
\affiliation{
  \institution{Tencent Inc.}
  \country{}
}
\email{octopuschen@tencent.com}

\author{Peng Chen}
\affiliation{
  \institution{Tencent Inc.}
  \country{}
}
\email{pengchen@tencent.com}

\author{Guoliang Li}
\affiliation{
  \institution{Tsinghua Univ.}
  \country{}
}
\email{liguoliang@tsinghua.edu.cn}

\author{Xuanhe Zhou}
\affiliation{
  \institution{Shanghai Jiao Tong Univ.}
  \country{}
}
\email{zhouxuanhe@sjtu.edu.cn}

\author{Fan Wu}
\affiliation{
  \institution{Shanghai Jiao Tong Univ.}
  \country{}
}
\email{fwu@cs.sjtu.edu.cn}

\begin{abstract}
Vector search systems are essential infrastructure for modern data-driven applications. Large-$k$ analytical vector search, which retrieves $k=10^3$--$10^5$ results for analytics (e.g., aggregation, filtering, joins), is increasingly important for emerging workloads, including LLM data management and advertising analysis at Tencent. Existing systems remain inadequate: specialized vector databases often cap $k$ (e.g., $k \leq 10^4$) to satisfy tail-latency constraints and offer limited analytical support, while OLAP systems typically embed per-segment vector indexes as black boxes, causing severe read/compute amplification and preventing native query optimization.



This paper presents \ourSys, an OLAP-native vector search system for large-$k$ workloads. \ourSys makes vector search a first-class analytical primitive in Tencent's OLAP engine through a global segment-decoupled index materialized as relational tables, eliminating scatter--gather execution, reducing amplification, and enabling native storage optimizations. It decomposes IVFPQ-based search into relational operators, integrates OLAP optimizations, and introduces \ourPQ, which combines direction-aware quantization with hierarchical residual refinement to improve recall while preserving relational efficiency. \ourSys further incorporates index-aware query rewriting and a distributed-aware cost model for efficient distributed execution. Experiments show that \ourSys achieves up to a $145\times$ speedup over competitive systems such as StarRocks, and up to a $52\times$ improvement in 10-billion-scale production deployments. 

\end{abstract}

%% file: Sections/1-Introduction.tex
\begin{table}[!t]
\centering
\vspace{1em}
\caption{Comparison of small-$k$ and large-$k$ workloads.}
\vspace{-0.1in}
\label{tab:query-types}
\resizebox{\linewidth}{!}{
\begin{tabular}{ccccc}
\toprule
\textbf{Workload} & 
$\boldsymbol{k}$ & 
\textbf{Latency} & 
\textbf{Serving} & 
\textbf{Applications} \\
\midrule

Small-$k$ &
$1 \sim 1\text{K}$ &
Low & 
Answers & 
Image Retrieval, RAG \\

Large-$k$ &
$1\text{K} \sim 100\text{K}$ & 
Moderate & 
Exploratory & 
\makecell{Training Set Retrieval,\\Advertising Data Analysis} \\

\bottomrule
\end{tabular}}
\vspace{-0.2in}
\end{table}

\vspace{-1.5em}
\section{Introduction}
\label{sec:intro}
Vector search has become a fundamental building block of modern database and multimodal search systems~\cite{pgvector, elasticsearch, clickhouse, GaussDB-Vector, doris, starrocks, Singlestore-v, BlendHouse, vs4future}. It typically involves two common query types: Approximate $k$-Nearest Neighbor Search (ANNS), which retrieves the top-$k$ most similar vectors from a large-scale dataset, and Filtered ANNS (FANNS), which incorporates attribute-based predicates to restrict the search scope and returns the top-$k$ most similar vectors within the filtered subset.
With the emergence of new applications, vector search workloads are no longer confined to small-$k$ scenarios. Instead, many applications require a large number of results, typically ranging from $10^3$ to $10^5$. For example, in our production environment at Tencent, we maintain a corpus of over 10 billion images within TEngineDB-V to serve the training set retrieval and analysis requests for Text-to-Image/Visual model developers~\cite{ldm, dall_e, clip, imagen}. They typically retrieve $10^4$--$10^5$ images to build domain-specific fine-tuning datasets~\cite{dreambooth, datacomp}. 
They also use our system to compute aggregate statistics that characterize the quality and distribution of training sets that either semantically correspond to particular text prompts or are visually similar to given reference images. 

\begin{figure}[t]
    \centering
    \includegraphics[width=\linewidth]{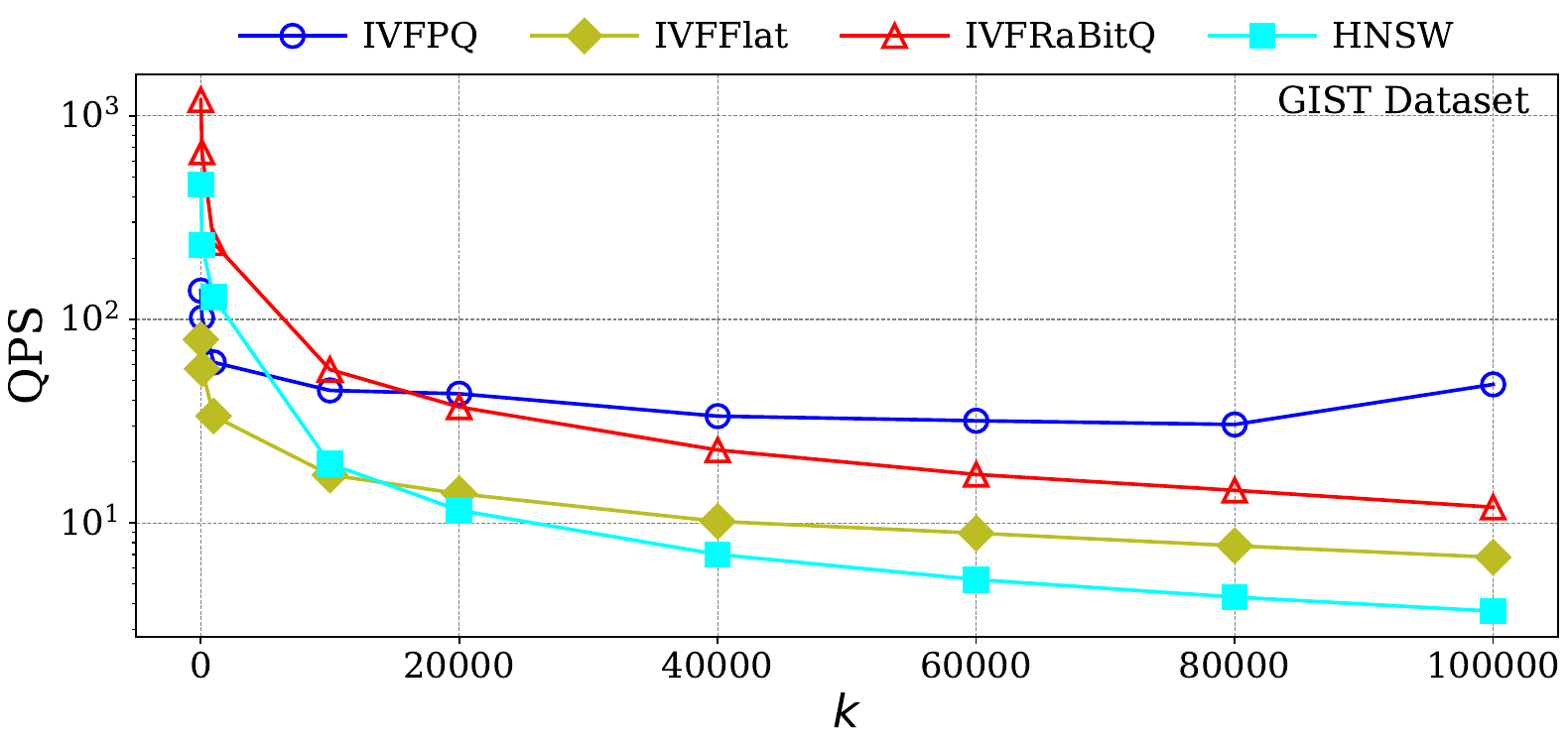}
    \vspace{-0.3in}
    \caption{QPS comparison of vector indexes under varying $k$.}
    \label{fig:pre-experiment}
\end{figure}

\begin{table*}[!t]
\centering
\caption{Comparison of representative specialized and database-integrated vector search systems. “Segment-coupled” indicates per-segment local indexing, whereas “Segment-decoupled” denotes a globally organized index across segments.} 
\label{tab:large-k}
\vspace{-0.15in}
\footnotesize
\setlength{\tabcolsep}{6pt}
\rowcolors{2}{gray!8}{white}
\resizebox{\linewidth}{!}{
\begin{tabular}{l c c l c c c}
\toprule
\textbf{System} 
& \textbf{Index Persistence} 
& \textbf{Index Organization} 
& \textbf{Index Type} 
& \textbf{Max $k$} 
& \textbf{Execution Operator} 
& \textbf{Native Integration} \\
\midrule
Milvus~\cite{Milvus} 
& Binary files 
& Segment-coupled 
& HNSW, Vamana, IVFPQ, IVFRaBitQ
& 16K (default) 
& Dedicated 
& Specialized \\

PGVector~\cite{pgvector} 
& Binary files 
& Segment-decoupled 
& HNSW, IVFFlat 
& 10K
& Dedicated 
& \xmark \\

GaussDB-Vector~\cite{GaussDB-Vector} 
& Binary files 
& Segment-coupled 
& Vamana, IVF-Based 
& Not disclosed 
& Dedicated 
& \xmark \\

ElasticSearch~\cite{elasticsearch} 
& Binary files 
& Segment-coupled 
& HNSW, BBQ 
& 10K 
& Dedicated 
& \xmark \\

Doris~\cite{doris} 
& Binary files 
& Segment-coupled 
& HNSW, IVF-Based 
& Unlimited 
& Dedicated 
& \xmark \\

StarRocks~\cite{starrocks} 
& Binary files 
& Segment-coupled 
& HNSW, IVFPQ 
& Unlimited  
& Dedicated 
& \xmark \\

SingleStore-V~\cite{Singlestore-v}
& Binary files 
& Segment-coupled 
& HNSW, IVFFlat, IVFPQ
& Not disclosed  
& Dedicated 
& \xmark \\ 

BlendHouse~\cite{BlendHouse}
& Binary files 
& Segment-coupled 
& HNSW, Vamana, IVF-Based
& Unlimited  
& Dedicated 
& \xmark \\ 

\textbf{\ourSys (ours)} 
& \textbf{Tables} 
& \textbf{Segment-decoupled} 
& \textbf{IVFPQ, \ourIVFPQ (ours)}
& \textbf{Unlimited} 
& \textbf{Relational} 
& \cmark \\

\bottomrule
\end{tabular}
}
\end{table*}

As shown in Table~\ref{tab:query-types}, unlike traditional small-$k$ workloads, which return only a few results under strict latency constraints~\cite{ann-benchmark}, large-$k$ workloads target exploratory analytics over substantial similarity sets. Their latency requirements are inherently more relaxed, as the computational cost grows significantly with $k$. This shift introduces new requirements to the underlying systems:
(1) \emph{Scalability under large result cardinality.} The system must efficiently process numerous results within acceptable latency, while avoiding excessive read/compute amplification. 
(2) \emph{Native integration with analytical execution.} Large-$k$ retrieval is rarely the final step. Instead, the retrieved similarity sets are further processed by downstream analytical operators (e.g., filtering, aggregation, join). Therefore, \textit{vector search should be tightly integrated into the analytical engine as a native operator}~\cite{ByteHouse} to enable joint optimization with other analytical operators.
However, existing vector search systems, including both specialized systems~\cite{Milvus, meta-faiss, SPTAG} and general-purpose database extensions~\cite{pgvector, GaussDB-Vector, elasticsearch, doris, starrocks, Singlestore-v, BlendHouse}, are primarily optimized for small-$k$ workloads and lack deep co-optimization between vector search operators and the native analytical pipelines. Therefore, they exhibit the following two main limitations.

\noindent
\textbf{Limitation 1: Limited Scalability to Large-$k$ Serving.} 
As shown in Table~\ref{tab:large-k}, many representative vector search systems~\cite{Milvus,pgvector,elasticsearch} 
impose explicit upper bounds on $k$ (typically 100–16K) to control tail latency. Although some systems (e.g., StarRocks~\cite{starrocks}, Doris~\cite{doris}) do not enforce a hard limit, they exhibit poor scalability when extended to large-$k$ retrieval, primarily due to two factors.
(1) \emph{Segment-coupled indexing and scatter–gather execution.}
Many systems~\cite{Milvus, BlendHouse, ByteHouse, doris, starrocks} maintain independent vector indexes per data segment and employ scatter–gather execution for top-$k$ vector similarity search.
When a table is split into $N$ segments, this forces the system to fetch at least $k$ records from each segment, inflating the result volume to 
$N \times k$ (e.g., 200M records for $k$=100,000 and $N$=2,000). This amplification cascades into 
excessive disk I/O, network traffic, and global merge costs.
(2) \emph{Degradation of graph-based search under large-$k$.}
Most systems rely on graph-based indexes (e.g., HNSW~\cite{HNSW}) for efficient small-$k$ retrieval. However, as $k$ increases, the candidate set expands rapidly, causing substantial computation for both vector evaluation and intermediate result maintenance (see details in Figure~\ref{fig:pre-experiment}).


\noindent \textbf{Limitation 2: Loose Integration with Underlying Systems.}
To support vector search and analytics, popular OLAP systems 
typically integrate vector indexes as a third-party component (e.g., using the Faiss~\cite{meta-faiss}  library) that is loosely coupled with the system~\cite{doris,starrocks,ByteHouse}. These indexes are stored as isolated files, and their execution logic remains black-box to the optimizer and compute engine. 
Therefore, general-purpose database optimizations cannot be directly applied to the index search. 
For example, systems typically invoke a black-box \texttt{IVFScan} operator to probe top-$n$ clusters, making the following general-purpose database optimizations, including cardinality estimation (lack visibility into cluster distributions), I/O pruning (fail to access specific clusters on disk selectively), and fine-grained caching (cache granularity in index file level rather than individual clusters), infeasible.


\begin{figure}[t]
    \centering
    \includegraphics[width=\linewidth]{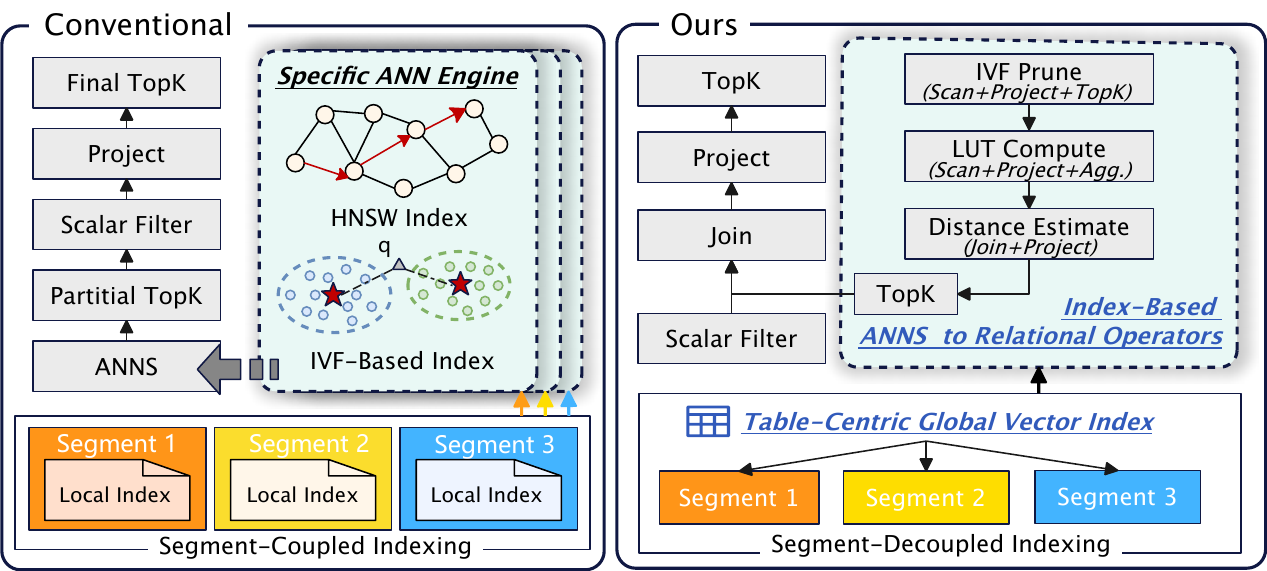}
    \vspace{-0.2in}
    \caption{Relational vector search paradigm.}
    \label{fig:paradigm}
\end{figure}

To this end, we propose \ourSys, an OLAP-native large-$k$ vector search engine built on the production-grade OLAP system \texttt{TEngineDB} deployed at Tencent. As illustrated in Figure~\ref{fig:paradigm}, \ourSys introduces a new paradigm for large-$k$ vector search: it materializes index structures as relational tables and executes vector search through composable relational operators. 

Specifically, (1) at storage layer, \ourSys builds a segment-decoupled global index spanning all data segments, which is fully decoupled from raw data and materialized as standard relational tables, enabling native optimizations such as columnar compression and fine-grained caching. 
(2) At the computation layer, \ourSys adopts a relational execution paradigm, transforming IVFPQ-based search into a composable sequence of relational operators and enabling native optimizations such as late materialization and runtime filters. This design is particularly effective for large-$k$ workloads, as it leverages the multi-threaded OLAP engine to parallelize key operations (e.g., inverted-list scanning, LUT construction, and distance computation), avoiding the scalability limitations of traditional single-threaded execution. However, traditional IVFPQ-based search often relies on low-precision distance estimation and therefore requires refining $2k$--$3k$ candidate vectors to ensure accuracy, making refinement increasingly expensive as $k$ grows. To address this issue, we develop \ourPQ, which improves distance estimation accuracy by redesigning the quantization mechanism and integrating refinement directly into fused relational operators, avoiding the costly large-scale raw-vector refinement stage. 
(3) At the control layer, we design an index-aware query rewriting module that injects vector index semantics into the relational optimizer and enumerates logically equivalent plans. A distributed-aware cost model further evaluates these plans by jointly modeling CPU, memory, and network costs. For large-$k$ hybrid queries, this optimization is essential since substantial intermediate results amplify join, materialization, and network overheads.


\noindent
\textbf{Product Impact.} 
\ourSys has been widely deployed in production environments at Tencent, powering advertising analysis, LLM services, and multimodal data analytics. Our largest dedicated cluster for large-$k$ workloads comprises 30+ compute nodes managing 100TB of data with about 10 billion image embeddings, consistently delivering <5s latency for top-100,000 search.


\hi{Contributions.} Our contributions are summarized as follows.

\begin{enumerate}[leftmargin=*, itemsep=0pt, labelindent=0pt,
  itemindent=0pt]
\item \textbf{We propose \ourSys, the first OLAP-native vector search system designed for large-$k$ workloads.} It introduces a new paradigm that tightly integrates IVFPQ-style indexing with relational storage and execution, enabling scalable large-result retrieval within production-grade OLAP infrastructures.

\item \textbf{We design a segment-decoupled global indexing framework with table-centric materialization.} These designs eliminate exhaustive scatter-gather execution across 
all segments and substantially reduce read/compute amplification under large-$k$ workloads, while enabling native storage-level optimizations.

\item \textbf{We present a relational execution paradigm for vector search with an enhanced quantization method.} 
We decompose IVFPQ-based search into composable relational operators, allowing seamless integration and OLAP-native optimizations. To reduce quantization error, we develop \ourPQ, which leverages direction-aware quantization and hierarchical residual refinement to improve accuracy while keeping relational efficiency.

\item \textbf{We develop an index-aware distributed optimization framework.} 
The framework injects vector index semantics into logical plans and jointly models CPU, memory, and network costs to select efficient execution plans in distributed environments.

\item \textbf{We conduct extensive experimental and production evaluations at billion scale.}
Experimental results demonstrate up to a $145\times$ speedup over competitive systems under large-$k$ workloads, while 10-billion-scale production deployment achieves up to a $52\times$ performance gain.
\end{enumerate}

%% file: Sections/2-Preliminaries.tex
\section{Preliminaries}
\label{sec: preliminary}
This section first presents two typical vector search query types, followed by an overview of representative vector search systems and the IVFPQ index.

\subsection{Two Query Types for Vector Search}
\textbf{ANNS.} 
Given a dataset $X = \{\mathbf{x}_1, \dots, \mathbf{x}_n\} \subset \mathbb{R}^d$ and a query vector $\mathbf{q} \in \mathbb{R}^d$, $k$-nearest neighbor search ($k$NNS) returns a subset $S_k \subset X$ with $|S_k| = k$ such that for any $\mathbf{x} \in S_k$ and $\mathbf{y} \in X \setminus S_k$, $\mathrm{dist}(\mathbf{q}, \mathbf{x}) \le \mathrm{dist}(\mathbf{q}, \mathbf{y})$. 
In high-dimensional spaces, exact $k$NNS is computationally prohibitive due to the curse of dimensionality~\cite{HNSW}. 
Approximate nearest neighbor search (ANNS) relaxes the accuracy requirement by returning vectors whose distances are close to those of the true top-$k$ neighbors with high probability, thereby substantially reducing computational overhead while maintaining strong empirical accuracy~\cite{kANNSurvey}.

\noindent
\textbf{FANNS.} Filtered Approximate Nearest Neighbor Search (FANNS) augments ANNS with attribute predicates evaluated alongside vector similarity~\cite{Mesh, liang2024unify, ADBV, SeRF}. Formally, given a filtering predicate $P: X \rightarrow \{\text{true}, \text{false}\}$, let $X_P = \{ \mathbf{x} \in X \mid P(\mathbf{x}) = \text{true} \}$ be the subset of vectors that satisfy the predicate. FANNS identifies a subset $S_k \subset X_P$ of size $\min(k, |X_P|)$ such that $\text{dist}(\mathbf{q}, \mathbf{x}) \le \text{dist}(\mathbf{q}, \mathbf{y})$ for any $\mathbf{x} \in S_k$ and $\mathbf{y} \in X_P \setminus S_k$.


\subsection{Existing Vector Search Systems}
\noindent
\textbf{Specialized Vector Search Systems and Libraries.} 
Milvus~\cite{Milvus}, Pinecone~\cite{pinecone}, Weaviate~\cite{weaviate}, Qdrant~\cite{qdrant}, FAISS~\cite{meta-faiss}, and SPTAG~\cite{SPTAG} are specialized ANNS systems or libraries built on PQ- or graph-based indexes (e.g., HNSW~\cite{HNSW}), primarily optimized for latency-sensitive small-$k$ workloads. As $k$ grows, the number of maintained candidates and intermediate results increases substantially, enlarging the effective search space and diminishing pruning effectiveness. Therefore, under large-$k$ analytical workloads, they often suffer from significantly reduced query efficiency.

\noindent
\textbf{Database-Integrated Vector Search Systems.} To bridge vector search and relational analytics, recent database and analytical systems have introduced vector search extensions within their architectures, including Elastic Search~\cite{elasticsearch}, PGVector~\cite{pgvector}, GaussDB-vector~\cite{GaussDB-Vector}, Doris~\cite{doris}, StarRocks~\cite{starrocks}, SingleStore-V~\cite{Singlestore-v}, and BlendHouse~\cite{BlendHouse}. 
In these designs, vector search is implemented as a specialized module whose execution logic remains opaque to the optimizer and compute engine. 
As a result, despite the availability of mature OLAP optimizations, e.g., columnar compression and vectorized execution, these modules cannot fully exploit them.

Under large-$k$ workloads, these systems often exhibit severe performance degradation. This is largely due to segment-coupled indexing and scatter–gather execution: each data segment maintains an independent local index, and execution requires probing all segments and merging their partial results. As $k$ increases, this fragmented execution amplifies I/O, intermediate materialization, and network overhead, leading to serious read/compute amplification.


\subsection{IVFPQ Algorithms}
IVFPQ~\cite{pq} is one of the most widely adopted indexing structures for large-scale ANNS. As our system is built upon an IVFPQ-based search, we briefly introduce its fundamental principles below.

IVFPQ uses $k$-means clustering~\cite{K-means} to partition vectors into $k$ clusters and searches the top-$N_{probe}$ clusters closest to the query.
To further accelerate distance computation within each cluster, Product Quantization (PQ) is applied to encode the vectors. 
The $d$-dimensional vector is partitioned into $m$ disjoint subspaces, each of dimension $d/m$. For each subspace, a separate codebook with $C$ centroids is learned using k-means clustering. Each vector is therefore represented by an $m$-dimensional code, where each dimension is an integer in $\{1, \dots, C\}$ indicating the assigned subspace centroid.

During query execution, the query vector $q$ is decomposed into $m$ sub-vectors $q_i$. For each subspace, a lookup table (LUT) is constructed by computing the squared distances between $q_i$ and all $C$ centroids. The approximate squared distance between $q$ and a vector $x$ is then computed as $\text{dist}(q, x)^2 = \sum_{i=1}^{m} d(j,i)$,
where $j$ is the PQ code index of $x$ in the $i$-th subspace, and $d(j,i)$ is the corresponding entry in the precomputed distance table. This formulation allows efficient evaluation using only $mC$ $(d/m)$-dimensional distance computations to construct the LUT and $mn$ table lookups to score $n$ candidate vectors. For ANNS, the vectors with the $k$ smallest distances are returned.

%% file: Sections/3-Overview.tex
\section{System Overview}
\label{sec:overview}

\subsection{System Architecture}

\begin{figure}[!t]
    \centering
    \includegraphics[width=.95\linewidth]{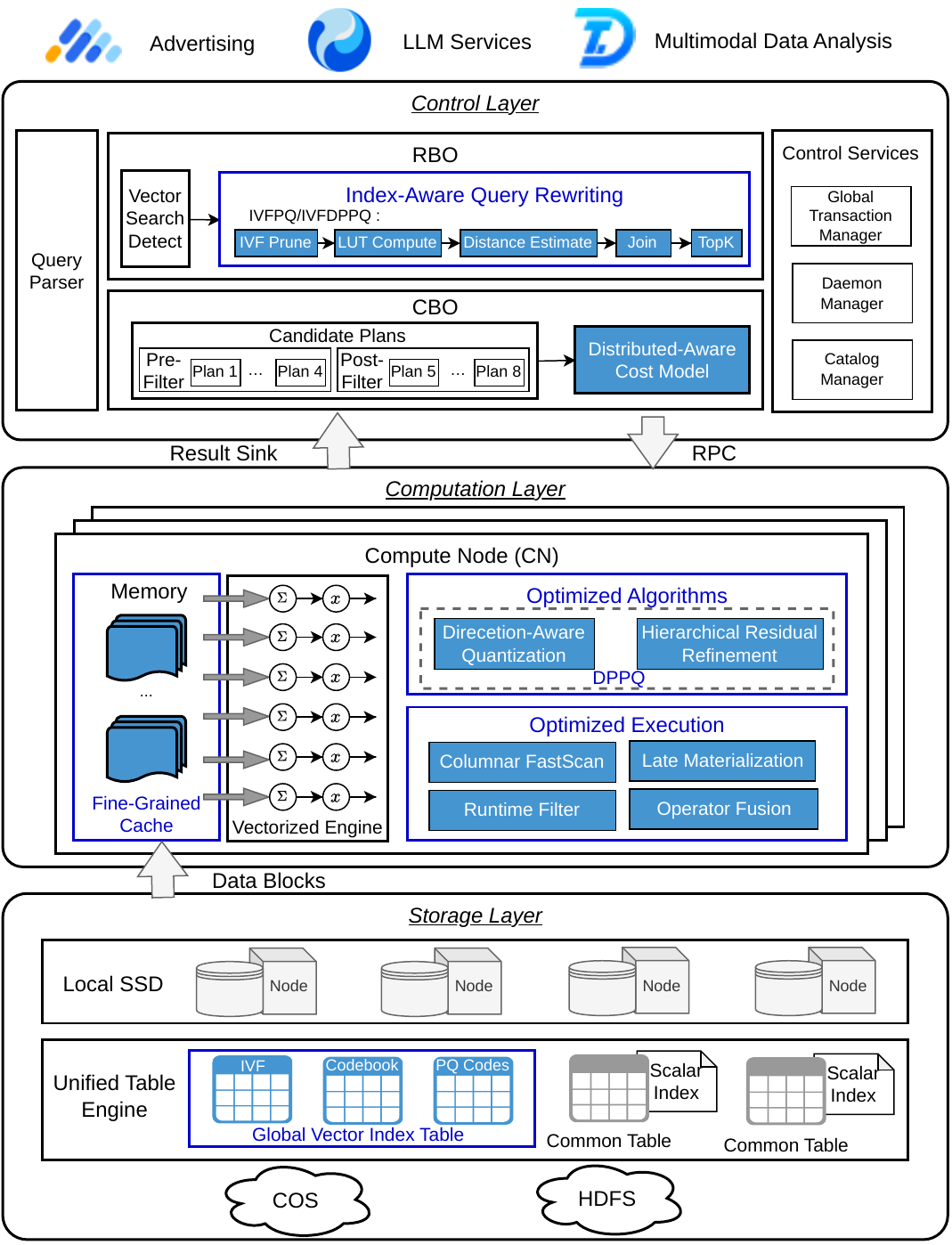}
   \vspace{-0.1in}
    \caption{The Architecture of \ourSys.}
   \vspace{-0.2in}
    \label{fig:overview}
\end{figure}

\begin{sloppypar}
\autoref{fig:overview} presents the system architecture of \ourSys, a cloud-native multi-modal data warehouse adopting a disaggregated storage paradigm where control plane, compute nodes, and storage layer operate as independent services. 

\noindent\textbf{Control Layer.}
The control layer integrates cluster-level management services (transaction, daemon, and catalog managers). Furthermore, it parses incoming SQL statements and performs both rule-based and cost-based optimizations (RBO and CBO) to generate efficient execution plans. Specifically, the RBO module recognizes vector/hybrid queries and rewrites them into index-aware operators. Instead of treating vector index search as a black-box operation, \ourSys decomposes IVFPQ search into explicit relational operators.
These operators can be natively optimized by the underlying OLAP compute engine.
Second, for FANNS queries, the CBO module enumerates alternative physical plans in consideration of various execution strategies (i.e., pre-filtering vs.\ post-filtering, left join vs.\ right join, and broadcast join vs.\ shuffle join). A distributed-aware cost model is developed to jointly estimate CPU, memory, and network I/O costs based on  data distribution and statistics, selecting the lowest-cost plan. The finalized distributed physical plan is then dispatched to the computation layer  via RPC.

\noindent
\textbf{Computation Layer.} The computation layer executes distributed plans across a cluster of compute nodes. We materialize vector indexes as standard relational tables (see storage layer) to leverage native OLAP optimizations: 
(1) \emph{Late Materialization}, which separates similarity computation from payload retrieval via a join operation; (2) \emph{Runtime Filters}, which prune irrelevant data at the source to reduce I/O and network traffic; 
(3) \emph{Columnar FastScan}, which aligns FastScan with the engine’s columnar layout to maximize SIMD efficiency;
(4) \emph{Operator Fusion}, which merges several operators into a single SIMD-optimized physical operator. 
Building on this OLAP-native design, we further introduce \ourPQ, which integrates direction-aware quantization (preserves directional fidelity to stabilize similarity ranking) with hierarchical residual refinement (progressively captures finer-grained directional components to mitigate quantization error), to improve retrieval accuracy.

\noindent
\textbf{Storage Layer.} The storage layer adopts a three-tier architecture (i.e., {remote storage}, {on-SSD cache}, and {in-memory cache}) to provide scalable and low-latency access of data and vector indexes. Built upon a unified table engine, \ourSys materializes both base data and vector indexes as relational tables. Instead of maintaining per-segment index files, it employs a \emph{segment-decoupled global indexing}: a single global index is built over the entire dataset, and its structures (IVF centroids, PQ codebooks, and quantized vector codes) are stored as sharded relational tables. This table-centric design eliminates the exhaustive scatter-gather execution inherent in per-segment indexing. Moreover, 
we adopt a two-tier cache to improve remote storage access latency and throughput, where remote data files are logically split into fine-grained, fixed-size blocks and cached in both local SSD disk and memory.


\noindent\textbf{Workflow.} When a vector/hybrid SQL query arrives, the control layer parses the query and performs index-aware query rewriting and cost-based optimization to generate a distributed physical plan. The optimized plan is then dispatched to the computation layer, where compute nodes execute it using relational operators with various OLAP-native optimizations. During execution, required data and index segment pages are fetched from the storage layer through globally indexed, table-centric access. Finally, the vector search results are produced and returned through the result sink.
\end{sloppypar}

\subsection{SQL Dialects}
\label{sec:sql}

\lstdefinelanguage{SQL}{
  keywords={SELECT, FROM, WHERE, INSERT, UPDATE, DELETE, CREATE, TABLE, 
            INDEX, VIEW, JOIN, INNER, LEFT, RIGHT, ON, GROUP, BY, ORDER, 
            HAVING, COUNT, SUM, AVG, MAX, MIN, AS, AND, OR, NOT, NULL, bigint, varchar, array, float, GLOBAL, VECTOR, INTERVAL, day, USING, LIMIT, PROPERTIES, DISTRIBUTED},
  keywordstyle=\color{blue}\bfseries,
  comment=[l]{--},
  commentstyle=\color{gray}\itshape,
  string=[b]',
  stringstyle=\color{red},
  morestring=[b]",
  sensitive=true
}

\lstset{
    language=SQL,
    basicstyle=\ttfamily\small,
    numbers=left,
    numberstyle=\tiny\color{gray},
    stepnumber=1,
    numbersep=8pt,
    rulecolor=\color{gray!30},
    tabsize=2,
    breaklines=true,
    breakatwhitespace=true,
    showstringspaces=false,
    captionpos=b,
    abovecaptionskip=10pt
}

\ourSys offers a flexible, easy-to-use SQL interface for manipulating tables/indexes and performing vector search or complex analytics. For example, the following SQL creates a table that is hash-sharded into 512 buckets by the primary key id, with each row containing an embedding vector and several structured attributes.

\begin{lstlisting}
CREATE TABLE images (
    id bigint,
    embedding array<float>,
    category varchar,
    author_id bigint
)
PRIMARY KEY(id)
DISTRIBUTED BY HASH(id) BUCKETS 512;
\end{lstlisting}

The following statement builds a global IVFPQ index on \texttt{embedding} paired with primary key \texttt{id}.

\begin{lstlisting}
CREATE GLOBAL INDEX vector_index 
ON images(embedding,id) USING VECTOR
REFRESH IMMEDIATE ASYNC EVERY (INTERVAL 1 day)
PROPERTIES ("index_type"="ivfpq", "dim"="768", "nbits"="4", "nlist"="1024");
\end{lstlisting}

We adopt an asynchronous strategy for index refreshing: the system checks the base table at most once per day and triggers a full background rebuild upon detecting any changes (Section \ref{sec:storage-segment-decoupled}).

Finally, the following SQL demonstrates a complex query pattern frequently observed in our production workloads. It presents a hybrid query that combines structured filtering (e.g., high-resolution images with width > 1024) with vector search, and joins the retrieved visually similar images with campaign metadata to retain ad creatives from a specific industry for advertising analysis.

\begin{lstlisting}
SELECT t.image_id, t.category, t.dist
FROM (
   SELECT image_id, category, campaign_id,
          approx_l2_distance(embedding, [0.1,0.2,...]) AS dist
   FROM images WHERE width > 1024
   ORDER BY dist LIMIT 1000000
) t
JOIN campaigns c ON t.campaign_id = c.id
WHERE c.industry = 'automotive';
\end{lstlisting}

%% file: Sections/4-Storage.tex
\section{Unified Table Storage Engine}
\label{sec:storage}
In modern analytical databases and vector management systems~\cite{Milvus, GaussDB-Vector, elasticsearch, doris, starrocks, Singlestore-v, BlendHouse}, data is typically horizontally split into multiple segments based on scalar attributes (e.g., primary keys or timestamps), and vector indexes are constructed independently for each segment. These indexes are commonly stored as proprietary binary files (or sidecar files) tightly bound to their corresponding data segments. We refer to this design as \emph{segment-coupled indexing}. However, this design introduces two fundamental limitations.

\noindent
\textbf{(1) Segment-coupled indexing enforces a scatter-gather execution, leading to severe amplification.} 
When the dataset is sharded into thousands of segment-local indexes, query processing must broadcast the query to all segments, compute local top-$k$ candidates, and perform a centralized global merge. This architecture scales poorly as either $k$ or data volume increases. As they grow, both local search costs and global merging overhead expand significantly, amplifying overall query latency.

\noindent
\textbf{(2) Storing the vector index as an isolated file prevents the system from leveraging native optimizations.}
The index is maintained in a specialized format, which is incompatible with the standard data storage engine. As a result, general-purpose optimizations such as  columnar compression and fine-grained caching cannot be directly applied to index data. Instead, the compute engine accesses the index through rigid, black-box APIs, which limits fine-grained resource scheduling and operator fusion, and ultimately constrains execution efficiency.


Motivated by these, \ourSys adopts a unified storage architecture centered on two core designs: \emph{Segment-Decoupled Global Indexing}, which builds a global index to eliminate scatter–gather execution, and \emph{Table-Centric Index Storage}, which materializes indexes as relational tables to enable native optimizations. 
We first introduce the overall data organization and multi-tier caching mechanism in Section~\ref{sec:cache}, followed by detailed descriptions of these two designs in Sections~\ref{sec:storage-segment-decoupled} and~\ref{sec:storage-table-centric}.


\subsection{Data Organization and Multi-Tier Cache} 
\label{sec:cache}

\noindent\textbf{Table Data Organization.}
Following a similar storage design to popular OLAP systems Doris~\cite{doris} and StarRocks~\cite{starrocks}, \ourSys tables are horizontally partitioned (typically by time or range) to enable pruning. Each partition is then hash-bucketed into shards. One shard serves as the fundamental unit of physical data distribution and cluster scaling, mapping directly to an independent path on remote storage (COS/HDFS). Within each shard, rowsets represent immutable, atomic ingestion batches tagged with monotonic version numbers for MVCC; each rowset comprises one or more segments. A segment is a self-contained columnar file applying three optimization layers: flexible encoding (dictionary, bit-shuffle, RLE, etc.) at the column level, compression (LZ4, ZSTD, etc.) at the page level, and auxiliary indexes (ZoneMap, Bloom filters, etc.) at the file level to minimize I/O.

\noindent\textbf{Multi-Tier Fine-Grained Cache.} 
To mitigate remote storage access latency, each compute node maintains a two-tier, fine-grained caching system consisting of both in-memory and on-SSD caches. Raw data is logically split into fixed-size blocks (typically 1~MB) for fine-grained caching. A hierarchical lookup strategy is used for data access: queries first probe the memory tier, then fall back to the local SSD cache, and finally retrieve data from remote storage if necessary. When a block is evicted from memory, it is asynchronously demoted to the SSD tier; data evicted from the SSD tier is permanently discarded. Furthermore, \ourSys enables cross-node cache access during cluster scaling, allowing new nodes to fetch data from existing peers before resorting to remote storage, thereby eliminating cold-start penalties during elastic expansion.

\subsection{Segment-Decoupled Global Indexing}
\label{sec:storage-segment-decoupled}

As illustrated in Figure~\ref{fig:decoupledVScoupled}, rather than constructing independent local indexes for each data segment, we build a single global vector index that spans the entire dataset. This index is physically decoupled from the underlying data segments and stored as a separate table. This architecture is termed as \emph{segment-decoupled indexing}, which fundamentally eliminates the exhaustive scatter–gather execution pattern inherent in segment-coupled paradigm. Rather than broadcasting queries to thousands of segment-local indexes and merging their intermediate top-$k$ results, the query directly navigates the global index to identify relevant clusters and the search cost only scales with probed cluster count, significantly reducing read/compute amplification. 

However, constructing and refreshing large-scale global index is non-trivial, as it requires aggregating and clustering massive volumes of vectors across all data partitions and refresh the globally consolidated structure, imposing immense computational, network, and disk I/O overhead. We address these challenges as follows.

\noindent
\textbf{Global Index Construction.}
We leverage Apache Spark~\cite{Spark} to build the global index. To construct the index, we submit Spark jobs to read our data files and perform distributed clustering to generate the IVF and PQ metadata (e.g., centroids and codebooks). The system then associates each vector with an IVF cluster ID and encodes it into PQ codes, bulk-loading the resulting records into \ourSys's index tables. 
Using Spark offers distributed scalability for billion-scale vector clustering/encoding and fault-tolerant execution that ensures robust index construction.


\noindent
\textbf{Global Index Refreshing.}
In the segment-decoupled architecture, handling updates requires periodically rebuilding the global index, which is expensive. To avoid blocking online queries, we perform index refresh asynchronously via background Spark jobs and apply the new index through an atomic swap. As a result, \ourSys adopts an eventual consistency model: during the refresh interval, the index may lag behind the base table (e.g., deleted IDs may still appear, and newly inserted vectors are not visible until the next rebuild).
This design is well-suited for analytical workloads with infrequent updates (days/weeks). A similar approach can also be found in the materialized view refresh mechanisms of open-source OLAP systems like Doris~\cite{doris} and StarRocks~\cite{starrocks}, as well as in commercial products such as Snowflake~\cite{snowflake}.


\noindent
\textbf{Rethinking Global Index Storage.} 
While a unified global index eliminates scatter–gather overhead, it introduces a new systems challenge: how to efficiently materialize and access this large, consolidated index structure within storage engine. If the global index is stored as a single big opaque binary file, it would not only create I/O bottlenecks but also remain outside the scope of the engine’s native storage optimizations. This architectural tension directly motivates our second core design: \emph{Table-Centric Index Storage}.

\subsection{Table-Centric Index Storage}
\label{sec:storage-table-centric}
\ourSys adopts a table-centric index storage design, in which the global index is materialized as relational tables and sharded across segments instead of segment-coupled, opaque binary file. 

\begin{figure}[!t]
    \centering
    \includegraphics[width=\linewidth]{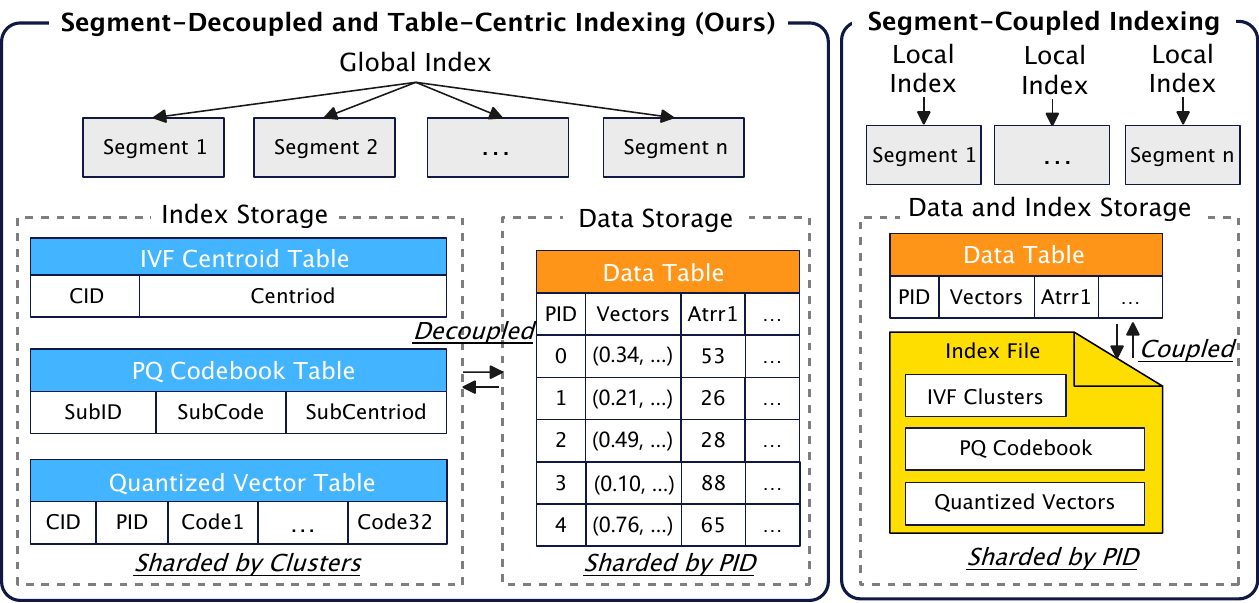}
    \vspace{-0.2in}
    \caption{Storage architectural comparison.}
    \vspace{-0.2in}
    \label{fig:decoupledVScoupled}
\end{figure}

\noindent
\textbf{Index Tables.} 
As illustrated in Figure~\ref{fig:decoupledVScoupled}, we decompose the IVFPQ index into three relational tables: 
(1) an \emph{IVF centroid table}, storing cluster identifiers (CIDs) and their centroids for cluster probing, sharded by CID; 
(2) a \emph{PQ codebook table}, maintaining subspace codebooks and not sharded; and 
(3) a \emph{quantized vector table}, mapping CIDs to vector identifiers (PIDs) and storing PQ codes, also sharded by CID.
PQ codes from different subspaces are materialized as separate columns to align with the columnar execution model and enable SIMD-friendly computation. 
Sharding by CID further ensures that only relevant shards are accessed during IVF probing.

\noindent
\textbf{Native Optimizations.}
Our table-centric design enables the global index to fully leverage native engine optimizations along the storage and execution pipeline.
(1) \uline{Columnar Compression.}
Since index tables are stored in columnar format, they automatically benefit from built-in compression schemes (e.g., dictionary encoding and run-length encoding). This reduces storage footprint and disk I/O without introducing custom, index-specific compression logic.
(2) \uline{Fine-Grained Cache.}
Rather than loading an entire index file into memory, \ourSys reads and caches only the required shards and blocks during query processing. This fine-grained caching mechanism reduces unnecessary I/O and improves memory efficiency.
(3) \uline{Vectorized Execution.}
Since index data is managed as relational tables, IVF filtering, distance computation, and top-$k$ selection can be executed using the engine’s vectorized operators, improving compute efficiency and CPU cache locality.

\noindent
\textbf{Rethinking Query Execution.}
Materializing the vector index as relational tables fundamentally reshapes the execution paradigm. Rather than invoking a specialized index interface, vector search is expressed through standard relational operators, enabling cluster routing, distance computation, and relational filtering to be jointly optimized by the SQL engine. 
This redesign raises three key challenges: 
(1) \textit{how to faithfully express and efficiently execute IVFPQ search within relational operators}; 
(2) \textit{how to mitigate the approximation errors inherent in PQ while preserving efficient relational execution}; and (3) \textit{how to fully exploit OLAP-native optimizations to accelerate search}. 
We address these challenges in the next section.



%% file: Sections/5-Query.tex
\section{OLAP-Native Computation Layer}
\label{sec:exec}

\begin{figure*}[!t]
    \centering
    \includegraphics[width=.95\linewidth]{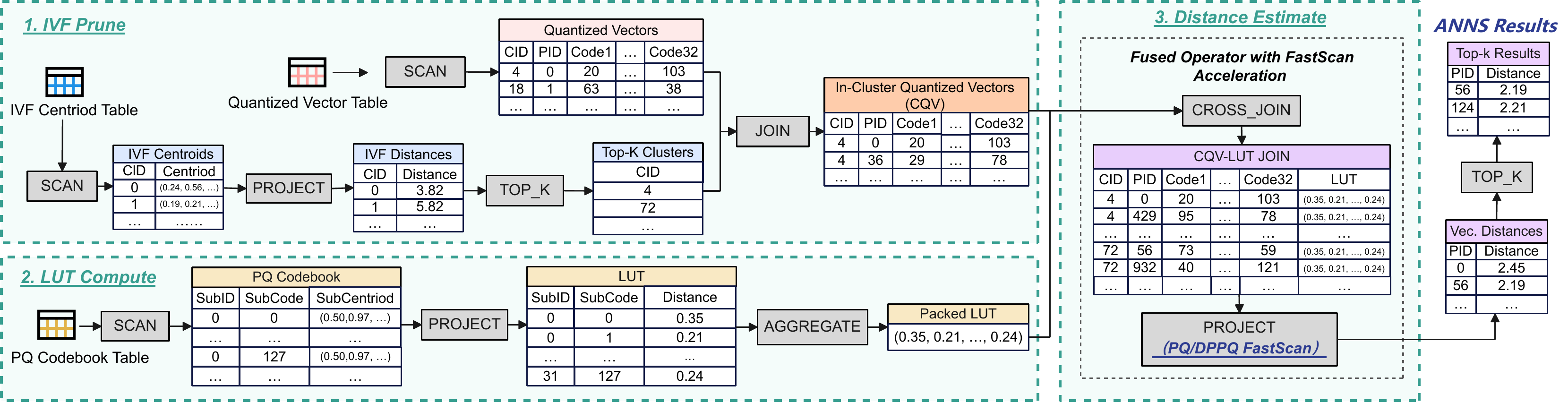}
    \vspace{-0.1in}
    \caption{Relational execution in \ourSys.}
    \vspace{-0.15in}
    \label{fig:exe}
\end{figure*}

To address the three challenges outlined above, this section presents the OLAP-native computation layer of \ourSys. Section~\ref{sec:operator} describes how classical IVFPQ search is adapted to the relational execution paradigm through operator-level decomposition. 
Section~\ref{sec:DPPQ} then improves retrieval accuracy within this framework by introducing \ourPQ. 
Finally, Section~\ref{sec:OLAP_OP} presents OLAP-native optimizations for relational execution.

\subsection{New Paradigm of Relational Execution}
\label{sec:operator}

As illustrated in Figure~\ref{fig:paradigm}, \ourSys executes IVFPQ-based vector search as a sequence of relational operators over the table-centric global index. Note (1) Graph-based methods~\cite{HNSW, NSG, tau-MNG, NHQ, TRIM, diskann, starling, SeRF, spann, rknn} are not considered as their poor performance under large-$k$ workloads (Figure~\ref{fig:pre-experiment}); (2) RaBitQ~\cite{RaBitQ, RaBitQ2} is another quantization approach with strong empirical performance, but its processing relies on complex operations, including per-cluster normalization, matrix multiplication, and bitwise operations, which are not naturally adaptable to standard relational operators and may be less efficient when adapted to relational execution.



\noindent
\textbf{Decomposing IVFPQ Search to Relational Operators.}
As illustrated in Figure~\ref{fig:exe}, IVFPQ search is mapped into three composable relational stages:
(1) \uline{IVF Prune.}
A \texttt{SCAN} over the IVF centroid table is first performed, followed by \texttt{PROJECT} to compute cluster–query distances and \texttt{Top\_K} to identify the $N_{probe}$ nearest clusters.
A subsequent \texttt{JOIN} between the selected clusters and the quantized vectors is performed to retrieve the quantized vectors within these clusters.
(2) \uline{LUT Compute.}
A \texttt{SCAN} over the PQ codebook table is executed, and \texttt{PROJECT} constructs the query-specific LUT. \texttt{AGGREGATE} then converts these multi-row results into a single-row array to avoid memory replication during distance evaluation.
(3) \uline{Distance Estimate.}
A \texttt{CROSS\_JOIN} broadcasts the LUT to candidate vectors, and \texttt{PROJECT} looks up and accumulates subspace distances to compute approximate distances. 
To further improve efficiency, we design a fused operator that consolidates these logical steps and accelerates execution using FastScan (Section~\ref{sec:OLAP_OP}). 
A final \texttt{Top\_K} selects the vectors with the smallest distances.




\subsection{\ourPQ: Bridging Accuracy and Relational Execution}
\label{sec:DPPQ}
Although IVFPQ can be efficiently executed within a relational framework, its achievable recall is fundamentally limited under large-$k$ workloads due to inherent quantization errors. 
A common remedy is to introduce a refinement phase that re-computes exact distances for an expanded candidate set (e.g., $2k$–$3k$ vectors with the smallest PQ distances). However, this approach requires materializing and re-evaluating high-dimensional raw vectors for a large candidate set, which quickly dominates query latency under large-$k$ workloads and negates the efficiency benefits of relational IVFPQ execution. Therefore, instead of relying on costly post-refinement, we enhance quantization accuracy at its source by redesigning the quantization mechanism and integrating refinement directly into the fused operator, avoiding large-scale raw-vector materialization.

To address this, we propose \ourPQ, a \textbf{\uline{D}}irectional \textbf{\uline{P}}rogressive \textbf{\uline{P}}roduct \textbf{\uline{Q}}uantization scheme that improves retrieval accuracy while remaining fully compatible with relational execution, eliminating the need for raw-vector materialization.

\noindent
\textbf{Direction-Aware Quantization.}
Unlike conventional PQ, which minimizes the mean squared reconstruction error $|\mathbf{x}-\hat{\mathbf{x}}|^2$ (where $\hat{\mathbf{x}}$ denotes the quantized vector), \ourPQ explicitly prioritizes directional alignment. Specifically, for a given vector $\mathbf{x}$, instead of directly minimizing Euclidean reconstruction error, \ourPQ minimizes the angular deviation between the normalized vector $\frac{\mathbf{x}}{|\mathbf{x}|}$ and its quantized representation. The intuition is that nearest neighbor ranking is often more sensitive to directional similarity than to absolute magnitude differences. Concretely, a high-dimensional vector $\mathbf{x} \in \mathbb{R}^D$ is partitioned into $m$ sub-vectors,
$\mathbf{x} = [\mathbf{x}^{(1)}, \mathbf{x}^{(2)}, \dots, \mathbf{x}^{(m)}]$.
Each sub-vector is decomposed into a scalar norm and a unit direction:
\begin{equation}
    \mathbf{x}^{(j)} = \|\mathbf{x}^{(j)}\| \cdot \frac{\mathbf{x}^{(j)}}{\|\mathbf{x}^{(j)}\|} = \rho^{(j)} \cdot \mathbf{u}^{(j)}
\end{equation}
where $\rho^{(j)}$ denotes the Euclidean norm and $\mathbf{u}^{(j)}$ is the unit direction. The directional components $\mathbf{u}^{(j)}$ are encoded using a PQ-style subspace clustering scheme.
The corresponding magnitudes $\rho^{(j)}$ are retained to preserve scale information and enable accurate distance reconstruction. The Euclidean distance is then reconstructed via the law of cosines by combining the approximated dot product with the precomputed norms of the query and base vectors:
\begin{equation}
    \| \mathbf{q} - \mathbf{x} \|^2 \approx \| \mathbf{q} \|^2 + \| \mathbf{x} \|^2 - 2 \sum_{j=1}^{m} \rho^{(j)} \langle \mathbf{q}^{(j)}, \mathbf{u}^{(j)} \rangle.
\end{equation}

\noindent
\textbf{Hierarchical Residual Refinement.}
To further reduce quantization error, \ourPQ employs a hierarchical residual quantization scheme that progressively refines the directional component. Let $\mathbf{r}_0 = \mathbf{x}$ denote the original vector. At layer $l$ ($1 \le l \le L$), we quantize the residual $\mathbf{r}_{l-1}$ using a directional-aware quantizer $Q_l(\cdot)$ and compute the new residual as $\mathbf{r}_l = \mathbf{r}_{l-1} - Q_l(\mathbf{r}_{l-1})$. This procedure is applied recursively across $L$ layers. The final approximation is therefore given by $ \hat{\mathbf{x}} = \sum_{l=1}^{L} Q_l(\mathbf{r}_{l-1})$. By successively encoding finer residual components, this hierarchical refinement progressively reduces quantization error.

\noindent
\textbf{\ourIVFPQ Index Construction.}
Algorithm~\ref{alg:dppq} describes the construction of \ourIVFPQ, consisting of a coarse clustering phase followed by multi-epoch residual encoding. We first partition the base vectors using K-Means~\cite{K-means} to obtain coarse centroids and build the IVF structure. 
Then, over $E$ epochs, residual vectors are hierarchically quantized: each residual is split into $M$ sub-vectors, decomposed into unit directions and scalar magnitudes, and the directions are encoded using PQ codebooks trained per epoch. After each epoch, residuals are reconstructed and updated by subtracting the decoded approximations.

\noindent
\textbf{Decomposing \ourIVFPQ Search to Relational Operators.} 
\ourPQ is integrated with the IVF structure for query execution. 
As illustrated in Figure~\ref{fig:exe}, \ourIVFPQ follows the same relational pipeline as conventional IVFPQ. The key distinction lies in the fused operator, which incorporates additional execution logic to retrieve the associated $\rho^{(j)}$ values for each candidate vector to reconstruct the DPPQ distance. When residual refinement is enabled, the operator iteratively executes these lookup and calculation steps for $epoch$ rounds to progressively refine the estimated distance.

As demonstrated in Section~\ref{sec:exp-dppq}, under the same bit budget, \ourPQ achieves a higher recall upper bound than the state-of-the-art RaBitQ, highlighting its superior quantization effectiveness.

\vspace{-1em}
\subsection{OLAP-Native Optimizations}
\label{sec:OLAP_OP}
Building upon the table-centric storage and relational execution, we introduce additional optimizations that deeply integrate with the underlying OLAP engine to accelerate large-$k$ vector search.

\noindent \textbf{(1) Late Materialization.} 
We decouple vector retrieval from attribute fetching via a \texttt{JOIN} operator that forms a two-phase execution pipeline. In the first phase, the system searches the vector index to obtain the global top-$k$ vector IDs. In the second phase, these IDs are joined with the base table to fetch the corresponding payload columns. Since the base table is hash-bucketed by vector ID, this join is executed efficiently as a hash join without shuffling data of the base table across compute nodes. This design inherently establishes a late materialization paradigm, preventing non-vector payload columns from participating in the I/O and network transmission during the similarity search phase, which is particularly beneficial for wide OLAP tables with numerous attributes.

\noindent \textbf{(2) Join Runtime Filter.} 
Join runtime filters \cite{bloomFilter, bloomJoin} are used to further optimize the execution of hash join operators in \ourSys. During execution, \ourSys generates runtime filters (including SIMD-accelerated Bloom filters and Min/Max ranges) on the build side and pushes them down to probe-side operators, filtering data before I/O and network transmission. This proves particularly effective for late materialization in vector search, where selective top-$k$ IDs aggressively prune the base table, minimizing redundant row access overhead for vector search.

\noindent \textbf{(3) Top-K Runtime Filter.} 
During distributed execution, the merge coordinator maintains a dynamic threshold, determined by the current $k$-th highest similarity score among received partial results. As stronger candidates arrive, the threshold is updated and pushed to worker nodes, which use it to prune non-competitive vectors during local top-$k$ search. By filtering candidates at the source, this mechanism substantially reduces cross-node data transfer under large-$k$ workloads.


\noindent \textbf{(4) Columnar PQ FastScan.}
PQ distance computation is commonly accelerated by FastScan \cite{FastScan1}, which uses SIMD instructions to perform efficient distance lookups and accumulation. Conventional FastScan \cite{FastScan1, FastScan2} assumes a row-oriented layout where PQ codes of 32 consecutive vectors are interleaved (packed) to enable 256-bit SIMD (e.g., AVX2) operations. However, OLAP compute engines \cite{doris,starrocks,ByteHouse}, typically adopt an Arrow~\cite{apache_arrow}-like columnar storage layout, creating a fundamental mismatch with the horizontal, interleaved data organization required by traditional FastScan.

\begin{algorithm}[!t]
\caption{Hierarchical Quantization of \ourIVFPQ}
\label{alg:dppq}
\small 
\KwIn{Base vector set $V$, Cluster count $Nlist$, Epochs $E$, Sub-spaces $M$, Bits per subspace $b$}
\KwOut{\ourIVFPQ Index Data}

$K_{sub} \leftarrow 2^b$\;
$d_{sub} \leftarrow D / M$\;

\tcp{Step 1: Coarse KMeans Clustering}
$C \leftarrow \text{KMeansTrain}(V, K_{c})$

$A \leftarrow \text{KMeansAssign}(V, C)$

\tcp{Step 2: Multi-Epoch Training and Encoding}

\For{$e \leftarrow 1$ \KwTo $E$}{
    \tcp{Decompose sub-vectors}
    $u_e, \rho_e \leftarrow \text{NormalizeSubvectors}(V, M)$\;
    
    \tcp{Train PQ codebook for unit direction and encode}
    $CB_e \leftarrow \text{PQTrain}(u_e, M, b)$\;
    $P_e \leftarrow \text{PQEncode}(u_e, CB_e)$\;
    
    \tcp{Decode, denormalize and update residuals}
    $\hat{u_e} \leftarrow \text{PQDecode}(P_e, CB_e)$\;
    $\hat{u_e} \leftarrow \text{DenormalizeSubvectors}(\hat{u_e}, \rho_e, M)$\;
    $V \leftarrow V - \hat{R}^{(e)}$\; 
    $CB_{all} \leftarrow CB_{all} \cup \{CB^{(e)}\}$\;
    $P_{all} \leftarrow P_{all} \cup \{P^{(e)}\}$\;
    $\rho_{all} \leftarrow \rho_{all} \cup \{\rho_e\}$\;
}
\Return $C, CB_{all}, (A, P_{all}, \rho_{all})$\;
\end{algorithm}

To bridge this gap, we develop a Columnar FastScan mechanism that adapts the FastScan algorithm to a columnar data layout without sacrificing SIMD efficiency. Recall that by our table-centric design, instead of storing PQ codes in an interleaved row-wise format, we organize the quantized codes as independent columns for different sub-quantizers (subspaces). Furthermore, the compute engine processes data in Arrow-like columnar chunks of 4,096 rows as its basic processing unit. Therefore, within each chunk, PQ codes with $m$ sub-quantizers are organized as separate arrays (columns), each containing the quantized indices for 4,096 vectors.

Our SIMD-optimized implementation operates directly on this vertical layout: for each sub-quantizer column, the kernel loads 32 consecutive codes and performs batched lookups and accumulations using a precomputed 32-byte LUT. Empirically, this columnar design incurs only about 5\% overhead versus row-oriented FastScan, while achieving a $4\times$--$5\times$ speedup over non-FastScan execution, balancing OLAP compatibility with computational efficiency.


\noindent \textbf{(5) Operator Fusion.}
To reduce the overhead of decomposed relational execution, we design a fused physical operator for distance estimation. The operator combines four logical stages: (1) LUT row replication, (2) LUT-based distance projection, (3) norm-based scaling for \ourPQ, and (4) final summation. It first consumes the single-row LUT and caches it in memory as a local pipeline breaker, then streams the quantized vector table directly. During streaming, it processes each input chunk in one pass, using SIMD vectorization to perform LUT lookups, norm scaling, and distance accumulation against the cached LUT state. This single-pass execution avoids materializing massive intermediate results. Together with vectorized execution~\cite{monetdb} over columnar chunks, the fused operator improves CPU cache locality and reduces per-tuple computation overhead compared with traditional volcano-style execution~\cite{volcano}.


%% file: Sections/6-Plan.tex
\section{Index-Aware Distributed Optimization}

\begin{table*}[t]
\vspace{-.75em}
\centering
\small
\setlength{\tabcolsep}{5pt}
\renewcommand{\arraystretch}{1.1}
\caption{Notation used in the cost model (left) and analytical cost breakdown of the eight FANNS execution plans (right).}
\label{tab:costmodel}
\vspace{-0.15in}

\begin{minipage}[t]{0.18\linewidth}
\centering
\resizebox{0.95\linewidth}{!}{%
\begin{tabular}{@{}ll@{}}
\toprule
\textbf{Symbol} & \textbf{Meaning} \\
\midrule
$N$ & Total rows \\
$N_f$ & Rows after filter \\
$M_p$ & Rows after IVF Pruning \\
$K_{ann}$ & Post-filter candidates \\
$R_F$ & Scalar row width \\
$R_V$ & Vector row width \\
$C_{build}$ & Build cost \\
$C_{probe}$ & Probe cost \\
$C_{hash}$ & Hash cost \\
$C_{dist}$ & Distance cost \\
$P$ & \# Nodes \\
\bottomrule
\end{tabular}}
\end{minipage}
\hspace{0.02\linewidth}
\begin{minipage}[t]{0.75\linewidth}
\centering
\resizebox{0.99\linewidth}{!}{%
\begin{tabular}{@{}lccccc@{}}
\toprule
\textbf{Plan} 
& \multicolumn{2}{c}{\textbf{CPU Cost}} 
& \textbf{Memory} 
& \textbf{Network} \\
\cmidrule(lr){2-3}
& \textbf{ANNS} 
& \textbf{Join} 
&  &  \\
\midrule
Plan 1 (Pre-Filter-Broadcast-Probe) &
$M_p C_{dist}$ &
$M_p C_{build}+N_f C_{probe}$ &
$M_p R_V$ &
$P M_p R_V$ \\
Plan 2 (Pre-Filter-Shuffle-Probe) &
$M_p C_{dist}$ &
$(N_f+M_p)C_{hash}+M_p C_{build}+N_f C_{probe}$ &
$M_p R_V$ &
$N_f R_F+M_p R_V$ \\
Plan 3 (Pre-Filter-Broadcast-Build) &
$N_f C_{dist}$ &
$N_f C_{build}+M_p C_{probe}$ &
$N_f R_F$ &
$P N_f R_F$ \\
Plan 4 (Pre-Filter-Shuffle-Build) &
$N_f C_{dist}$ &
$(N_f+M_p)C_{hash}+N_f C_{build}+M_p C_{probe}$ &
$N_f R_F$ &
$M_p R_V+N_f R_F$ \\
Plan 5 (Post-Filter-Broadcast-Probe) &
$M_p C_{dist}$ &
$K_{ann}C_{build}+N_f C_{probe}$ &
$K_{ann}R_V$ &
$P K_{ann}R_V$ \\
Plan 6 (Post-Filter-Shuffle-Probe) &
$M_p C_{dist}$ &
$(N_f+K_{ann})C_{hash}+K_{ann}C_{build}+N_f C_{probe}$ &
$K_{ann}R_V$ &
$N_f R_F+K_{ann}R_V$ \\
Plan 7 (Post-Filter-Broadcast-Build) &
$M_p C_{dist}$ &
$N_f C_{build}+K_{ann}C_{probe}$ &
$N_f R_F$ &
$P N_f R_F$ \\
Plan 8 (Post-Filter-Shuffle-Build) &
$M_p C_{dist}$ &
$(K_{ann}+N_f)C_{hash}+N_f C_{build}+K_{ann}C_{probe}$ &
$N_f R_F$ &
$K_{ann}R_V+N_f R_F$ \\
\bottomrule
\end{tabular}%
}
\end{minipage}

\vspace{-0.1in}
\end{table*}

At control layer, the query optimizer of \ourSys builds upon the Cascades framework \cite{cascades, volcano, orca} and extends it to jointly reason about vector-index execution semantics and distributed plan generation. This enables execution plans that are both \emph{index-aware} and \emph{communication-efficient}, essential for large-$k$ vector search.

When a SQL query containing a vector predicate is submitted, it is first transformed into a logical operator tree and processed by the rule-based optimizer (RBO). The RBO performs deterministic, \emph{index-aware query rewriting} on the initial logical operator tree, injecting specialized vector-index execution logic to establish a baseline execution strategy. The rewritten plan is then inserted into the Memo of the cost-based optimizer (CBO) under the Cascades framework. During optimization, the CBO expands the Memo to explore equivalent logical formulations and candidate physical implementations. To evaluate these alternatives, we integrate a \emph{distributed-aware cost model} that jointly estimates CPU cost, memory pressure, network overhead, and recall impact. 
This enables the optimizer to select the globally optimal physical plan from the expanded search space. 

\vspace{-1em}

\subsection{Index-Aware Query Rewriting}
\label{sec:rewriting}

Upon receiving a high-level vector search request, \ourSys rewrites it into an index-aware logical plan and enumerates equivalent alternatives within the Cascades framework.

\noindent
\textbf{Logical Plans for ANNS.}
For pure ANNS, rewriting is triggered when the RBO detects a canonical pattern in the logical plan, i.e., a vector distance computation followed by a \texttt{Top\_K} operator. 
Upon matching this pattern, the RBO applies certain rules to rewrite the generic operators with specialized vector search operators that embed index semantics (Figure~\ref{fig:exe}). 
The rewritten plan is inserted into the corresponding equivalence classes in the Memo, allowing the CBO to evaluate it alongside standard relational alternatives.

\begin{figure}[t]
    \centering
    \includegraphics[width=\linewidth]{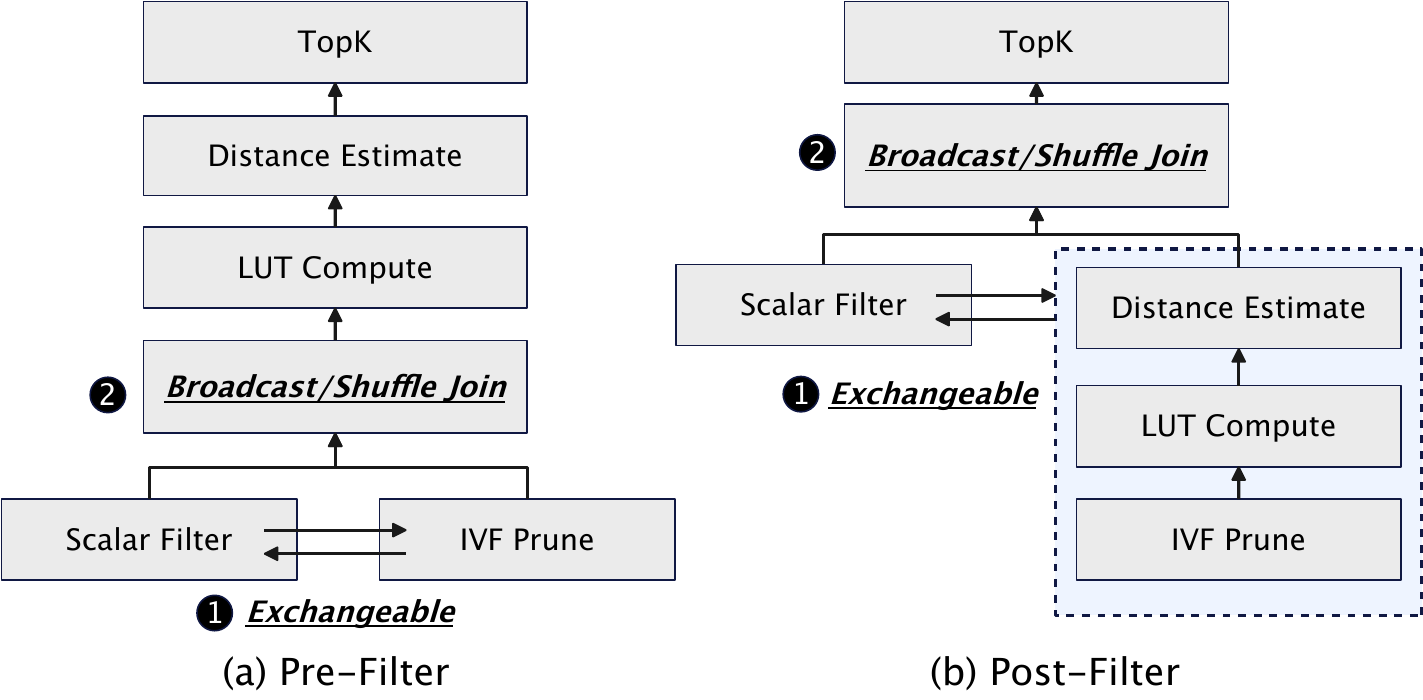}
    \vspace{-0.2in}
    \caption{Execution Plans for FANNS.}
    \vspace{-0.2in}
    \label{fig:executionPlans}
\end{figure}

\noindent
\textbf{Logical Plans for FANNS.} 
For FANNS queries that combine vector search with scalar predicates, a single static rewrite is insufficient, as the optimal strategy depends on both filter selectivity and data distribution. 
Our RBO first rewrites the query into a canonical \emph{pre-filter} plan by pushing scalar predicates below distance estimation. 
Once inserted into the Cascades Memo, transformation rules are triggered to explore the \emph{post-filter} alternative and other equivalent forms. As shown in Figure~\ref{fig:executionPlans}, we expand the search space along three orthogonal dimensions:
\uline{(1) Logical order.} \emph{Pre-filter} (scalar before distance) vs. \emph{post-filter} (scalar after distance).  
\uline{(2) Join order.} Two hash join configurations by swapping the build/probe side between vector and scalar intermediates.  
\uline{(3) Data exchange.} Two distributed strategies: \emph{broadcast} and \emph{shuffle} joins.

These three binary dimensions yield eight candidate physical plans 
($2 \times 2 \times 2$), all explicitly maintained in the Memo. This enumeration defines a complete execution search space, over which a \emph{distributed-aware cost model} is used to select the optimal plan.

\vspace{-.8em}
\subsection{Distributed-Aware Cost Model}
\label{sec:costmodel}
\vspace{-.4em}

Unlike existing systems that primarily model CPU cost~\cite{Milvus, ADBV, BlendHouse, GaussDB-Vector, pgvector}, we adopt a weighted cost model that jointly evaluates CPU, memory, network cost when evaluating alternative logical plans:
\begin{equation}
    Cost = w_\text{cpu} \cdot Cost_\text{cpu}
    + w_\text{mem} \cdot Cost_\text{mem}
    + w_\text{net} \cdot Cost_\text{net}.
\end{equation}
The weights are calibrated to reflect the architectural characteristics of \ourSys. By default, we set $w_\text{cpu}=0.5$, $w_\text{mem}=2.0$, and $w_\text{net}=1.5$, assigning higher penalties to memory pressure and network communication in distributed environments.


Recall that we classify the eight candidate FANNS plans along three orthogonal dimensions. For each dimension, we then analyze the corresponding CPU, memory, and network bottlenecks, with the overall cost breakdown summarized in Table~\ref{tab:costmodel}. 


\noindent
\textbf{Pre-Filter and Post-Filter.}
Pre-filtering reduces the vector search space to $N_F$ rows before distance evaluation, thus bounding the number of in-cluster candidates $M_p$ and ensuring recall correctness. 
However, when the scalar filter is unselective (i.e., $N_F \approx N$), both the join cost and memory usage become dominant, leading to CPU-bound execution. 
In contrast, post-filtering performs vector pruning first and joins only the top-$K_{ann}$ candidates, with CPU cost dominated by $M_p \cdot C_{dist}$ (i.e., distance computations on $M_p$ rows). 
Under highly selective scalar predicates, $K_{ann}$ must be set to a larger value to preserve recall, which increases join and memory costs and becomes inefficiency. Otherwise, insufficient setting of $K_{ann}$ may lead to recall degradation.

\ourSys adaptively sets $K_{ann}$ based on the predicate selectivity estimated by the optimizer (e.g., retrieving $2k$ candidates for an estimated 50\% selectivity), while also allowing manual overrides. Under the post-filtering strategy, instead of expanding the candidate pool using iterative search, the system does not strictly enforce returning exactly $k$ results. If the final filtered yield falls below $k$ but remains above a configurable threshold (defaulting to 20\% of $k$), it simply returns the available subset, as minor result reductions rarely impact the effectiveness of large-$k$ query workloads. However, if the yield drops below this threshold, the execution dynamically falls back to a pre-filtering plan.

\noindent
\textbf{Build-Side Selection.}
In hash join execution, the build side can be either the vector or scalar intermediate results. 
As shown in Table~\ref{tab:costmodel}, memory usage scales with the build-side cardinality and row width (e.g., $M_p \cdot R_V$, $K_{ann} \cdot R_V$, or $N_F \cdot R_F$), while CPU cost scales with the build size for hash construction and with the probe size for probing. 
Hence, assigning the smaller intermediate result to the build side is crucial for minimizing memory and construction overhead. 
When the vector candidate count is much smaller ($M_p \ll N_F$), building the hash table on the vector side is optimal. Conversely, when a selective scalar predicate results in a small filtered set ($N_F \ll M_p$), building on the scalar side is preferable. 

\noindent
\textbf{Broadcast Join and Shuffle Join.}
The network cost differentiates broadcast and shuffle joins. For broadcast joins (Plans 1, 3, 5, 7), the network cost scales linearly with the node size times the build-side size (e.g., $P \cdot M_p \cdot R_V$ or $P \cdot N_F \cdot R_F$). Therefore, broadcast is efficient only when the build side is sufficiently small. Otherwise, its cost grows proportionally with node scale and quickly dominates total latency.
In contrast, shuffle joins (Plans 2, 4, 6, 8) repartition both inputs across nodes, incurring a network cost proportional to the combined size of the intermediate results (e.g., $N_F \cdot R_F + M_p \cdot R_V$). 
Although shuffle avoids data replication and distributes memory pressure more evenly, it introduces additional hash partitioning overhead, such as $(N_F + M_p) \cdot C_{hash}$, cost to calculate hash codes for both vector search and scalar intermediate results. Consequently, shuffle joins become necessary when intermediate results are large or node scale amplifies broadcast cost. 

%% file: Sections/7-Experiments.tex
\section{Experiments}
\label{sec: experiments}

\subsection{Experiments Setting}
\label{sec:setting}

\textbf{Datasets.} 
We include low- to medium-dimensional image and text embeddings such as GloVe (100D), SIFT1M/SIFT1B (128D), and GIST (960D), as well as higher-dimensional textual embeddings from Wikipedia (768D). 
To assess scalability at extreme scale, we further incorporate Tencent-Image (768D), a real-world production dataset containing 10 billion vectors. 

\begin{table}[!t]
\centering
\caption{Statistics of dataset.}
\vspace{-0.15in}
\resizebox{\linewidth}{!}{
\begin{tabular}{ccccc}
\toprule
\textbf{Dataset} & 
\textbf{Dimension} & 
\textbf{\#Vectors} &
\textbf{\#Queries} &
\textbf{Source} \\
\midrule
GloVe & 100 & 1,183,514 & 10,000 & Texts\\
SIFT1M & 128 & 1,000,000 & 10,000 & Image\\
SIFT1B & 128 & 1,000,000,000 & 10,000 & Image\\
Wikipedia & 768 & 35,312,651 & 10,000 & Texts\\
Tencent-Image & 768 & 10,000,000,000 & 1,000 & Images\\ 
GIST & 960 & 1,000,000 & 1,000 & Images\\
\bottomrule
\end{tabular}}
\label{tab:statistics-of-dataset}
\vspace{-0.2in}
\end{table}


\noindent
\textbf{Compared Systems.} Both specialized and database-integrated vector search systems are compared. For specialized vector search systems, we use Milvus~\cite{Milvus}, a widely adopted and representative vector database. For database-integrated vector search systems, we evaluate StarRocks~\cite{starrocks} and PGVector~\cite{pgvector}. StarRocks is a distributed analytical database for real-time OLAP workloads. PGVector is a widely adopted PostgreSQL extension~\cite{PostgreSQL} that enables vector similarity search within relational databases. We incorporate the DiskANN library~\cite{diskann} to assess performance within the broader context of disk-resident indexing techniques~\cite{vs4future,diskresidentEA}. All evaluated systems are open-source and compared at medium recall levels (0.8–0.9), as in our target workloads vector search serves only as a coarse candidate generation stage rather than the final accuracy-critical results.


\noindent
\textbf{Parameter Settings.}
All systems are deployed using the same 56-shard configuration to ensure fair comparisons. All systems except DiskANN utilize the IVFPQ index. For the SIFT1B dataset, the data space is partitioned into 3000 clusters, and vectors are compressed using 32 subspaces with 8-bit quantization per subspace. For the Wikipedia dataset, the data is divided into 6000 clusters, utilizing 192 subspaces with 8-bit quantization. For DiskANN, the graph degree is set to 48, the build list size is 128, and the beam list parameter is 4 for both datasets. The search list size is dynamically adjusted, ranging from 36,000 to 58,500 for the SIFT1B dataset, and from 30,000 to 45,000 for the Wikipedia dataset.

\noindent
\textbf{Implementation.} Experiments are conducted on the machine equipped with an Intel Xeon Platinum 8576C processor (56 cores, 2.0 GHz), 487 GB of memory, and a 7 TB NVMe SSD. The case study is performed on a distributed cluster consisting of 10 machines, each with the exact same hardware configuration.

\begin{figure}[!t]
    \centering
    \includegraphics[width=\linewidth]{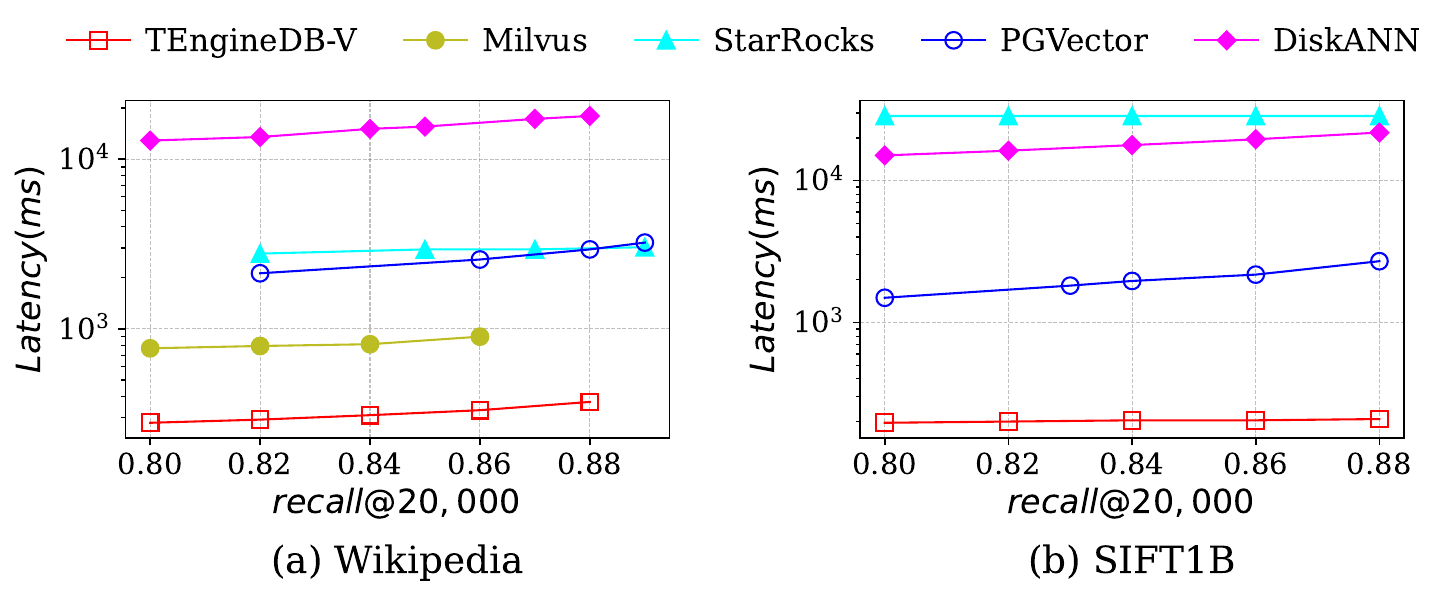}
    \vspace{-0.3in}
    \caption{Latency–Recall Trade-off Across Systems.}
    \vspace{-0.2in}
    \label{fig:overall-ANN}
\end{figure}

\subsection{ANNS Performance}
\label{sec:performance}

\begin{figure}[!t]
    \centering
    \includegraphics[width=\linewidth]{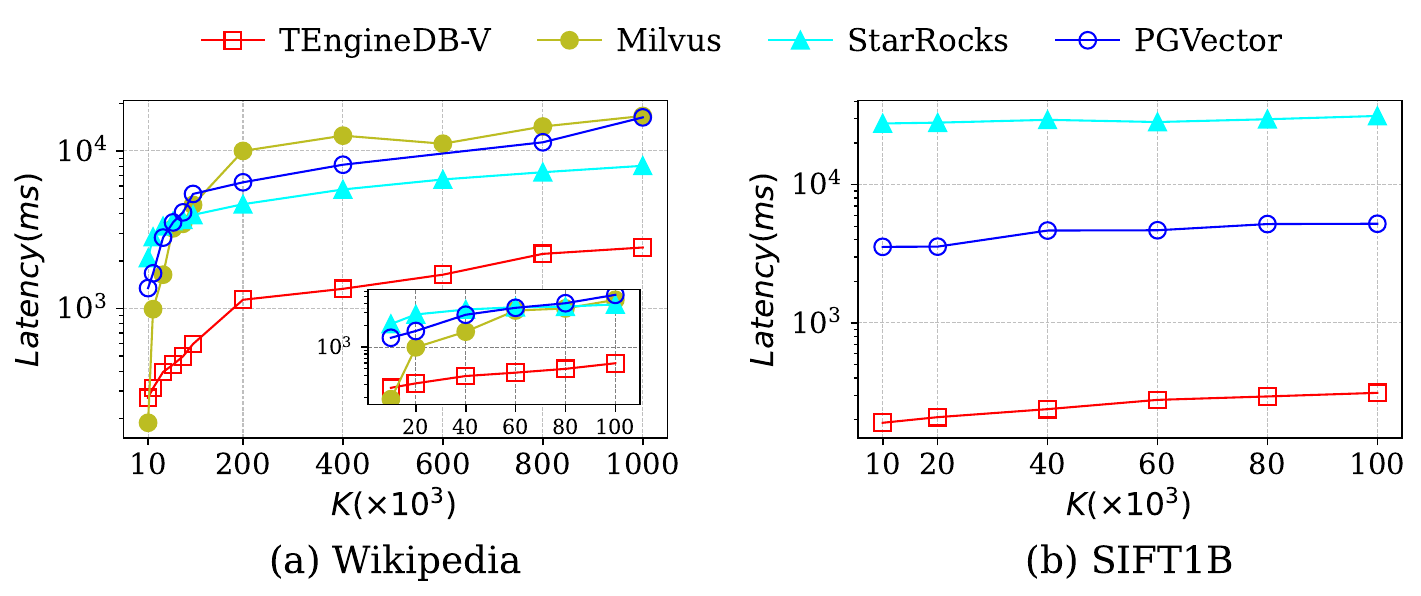}
    \vspace{-0.3in}
    \caption{Latency Scalability with Increasing $K$.}
    \label{fig:ANN diff k}
    \vspace{-1em}
\end{figure}

We compare the ANNS performance with $k = 20{,}000$. The results in \autoref{fig:overall-ANN} (Milvus does not report results on SIFT1B due to out-of-memory errors) show \ourSys consistently achieves superior latency–recall trade-off compared to all baselines. As shown in \autoref{fig:overall-ANN}(a), at a recall of 0.86, \ourSys achieves a latency of 332ms, representing speedups of $2.7\times$, $8.9\times$, $7.7\times$, and $50.0\times$ over Milvus, StarRocks, PGVector, and DiskANN library on the Wikipedia dataset. Similarly, in \autoref{fig:overall-ANN}(b), at a recall of 0.88, \ourSys achieves a latency of 208ms, representing speedups of $137.1\times$, $13\times$, and $104.7\times$ over StarRocks, PGVector, and DiskANN on SIFT1B. The performance gains mainly attribute to \ourSys’s segment-decoupled global indexing and relational execution framework: (1) the relational execution framework improves per-segment execution efficiency through multi-threaded parallel processing; (2) the segment-decoupled global index selectively probes only relevant segments, significantly reducing cross-segment intermediate-result merging and global top-$k$ sorting overhead. 



We also evaluate the performance under increasing $k$ values. On Wikipedia, $k$ ranges from $10^4$ to $10^6$, while on SIFT1B, $k$ ranges from $10^4$ to $10^5$, (\autoref{fig:ANN diff k}). Milvus does not report results on SIFT1B due to out-of-memory errors, as discussed earlier. Across both datasets, \ourSys consistently achieves lower latency than all baselines as $k$ increases. For example, when $k = 8 \times 10^5$, \ourSys achieves a latency of 2222ms, representing speedups of $6.4\times$, $3.3\times$, and $5.1\times$ over Milvus, StarRocks, and PGVector on Wikipedia, respectively. Similarly, when $k = 10^4$, it achieves a latency of 189ms, representing speedups of $145.5\times$ and $18.7\times$ over StarRocks and PGVector on SIFT1B. These results demonstrate the strong scalability of \ourSys with respect to increasing $k$, validating the effectiveness of OLAP-native ANNS in large-$k$ workloads.


\begin{figure}[!t]
    \centering
    \includegraphics[width=\linewidth]{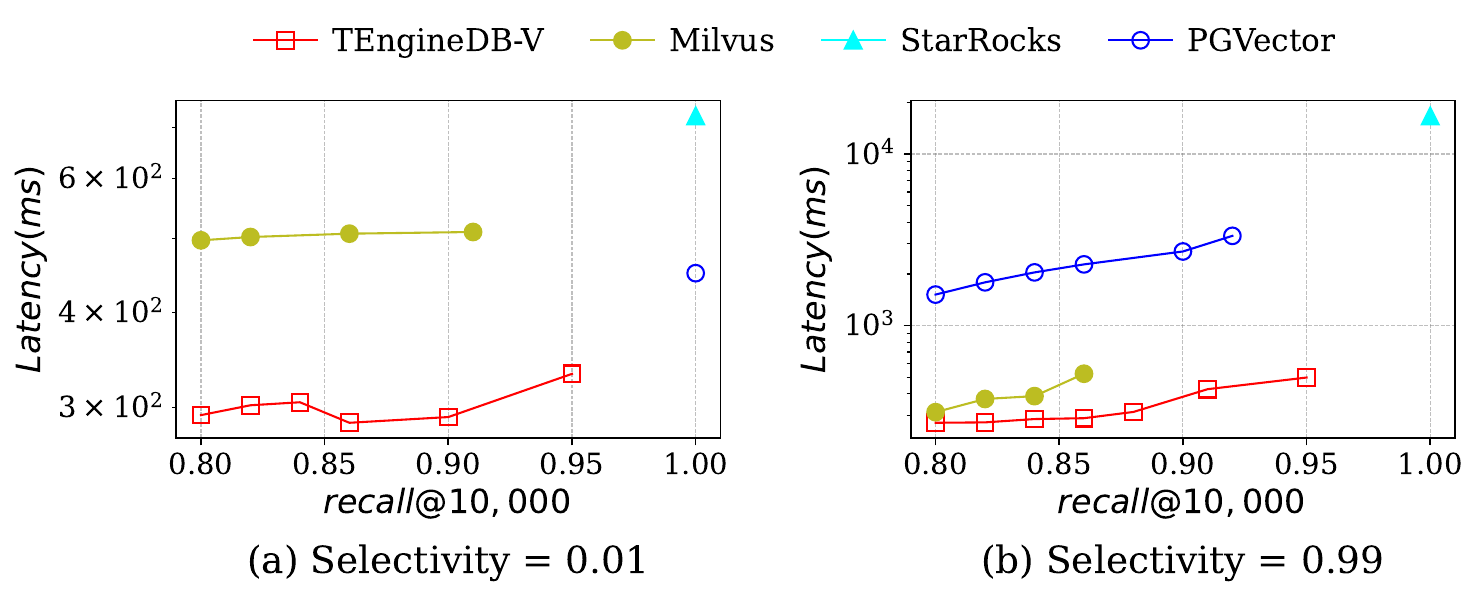}
    \vspace{-0.3in}
    \caption{FANNS Latency–Recall Trade-off on Wikipedia.}
    \vspace{-0.2in}
    \label{fig:overall-FANN}
\end{figure}

\begin{figure}[!t]
    \centering
    \includegraphics[width=\linewidth]{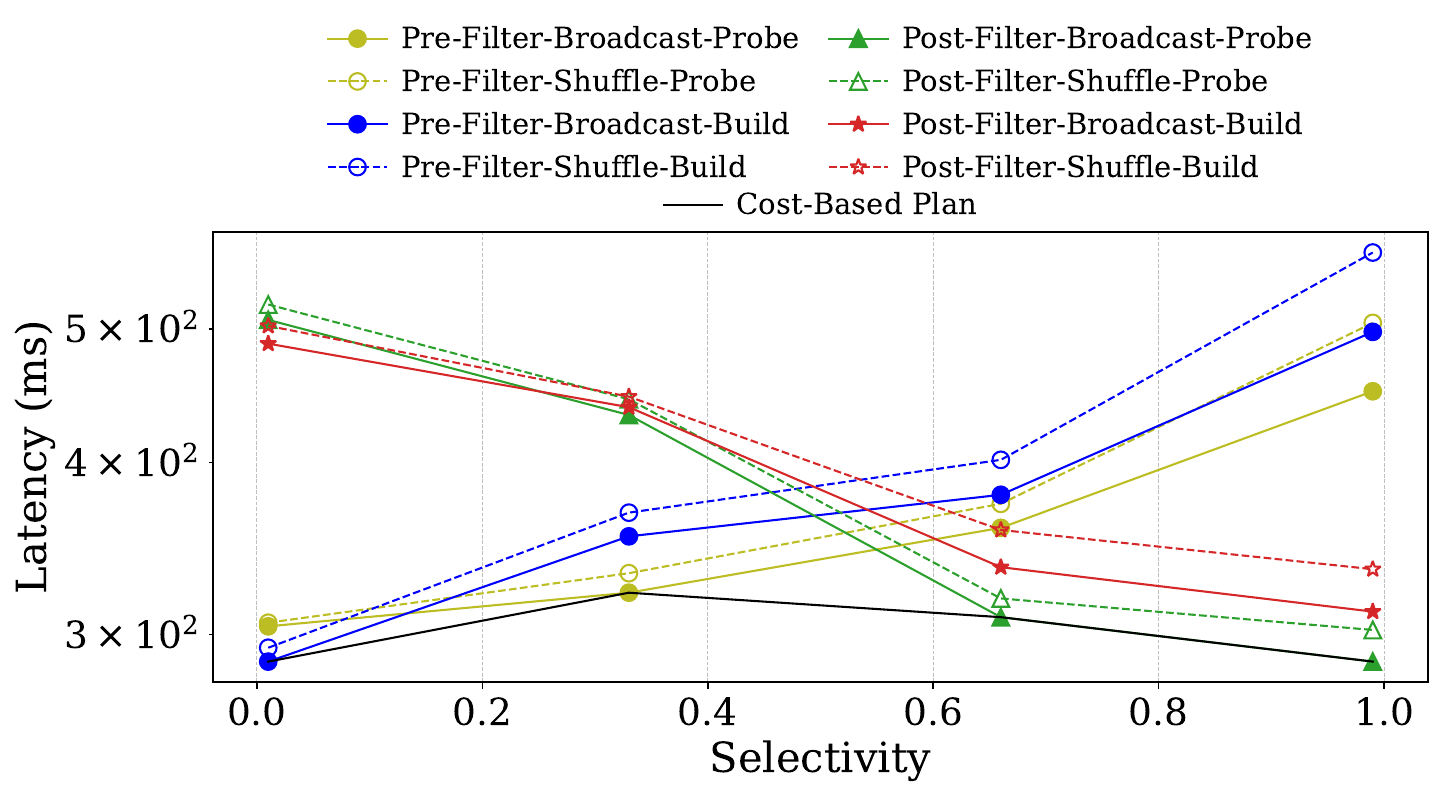}
    \vspace{-0.25in}
    \caption{Performance of the eight FANNS execution plans and the cost-based plan under varying selectivity.}
    \label{fig:cost_model}
    \vspace{-2em}
\end{figure}

\subsection{FANNS Performance}

We evaluate FANNS performance on the Wikipedia dataset under predicate selectivities of 1\% and 99\%. As shown in \autoref{fig:overall-FANN}, \ourSys consistently delivers the best latency-recall trade-off across selectivity settings. Under 1\% selectivity, \ourSys outperforms Milvus by $1.7\times$ at about 0.9 recall; under 99\% selectivity, it outperforms Milvus and PGVector by $1.6\times$ and $7.5\times$, respectively, at about 0.85 recall. These gains come from \ourSys's cost-based, distributed-aware FANNS optimizer, which selects efficient execution plans under different filtering conditions. We further validate effectiveness of the cost-based strategy with eight alternative plans across selectivities from 1\% to 99\% (\autoref{fig:cost_model}). Since no single plan dominates across all selectivities, the cost model achieves consistently strong performance by dynamically selecting the most suitable plan for each workload.

\begin{figure}[!t]
    \centering
    \includegraphics[width=\linewidth]{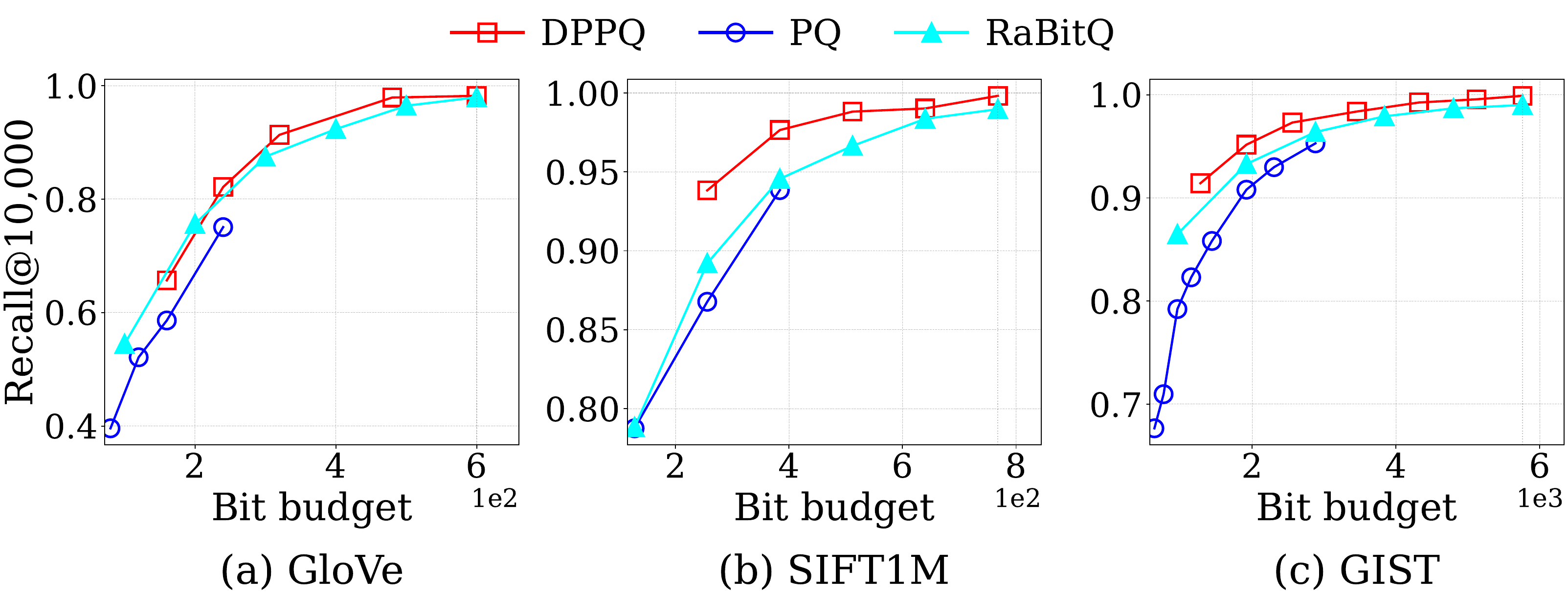}
    \vspace{-0.3in}
    \caption{Comparison of different quantization algorithms.}
    \vspace{-0.3in}
    \label{fig:Quantization_algorithm_comparison}
\end{figure}

\subsection{In-Depth Analysis of \ourPQ}
\label{sec:exp-dppq}
\subsubsection{Quantization performance.} 
We compare the quantization performance of \ourPQ against PQ and RaBitQ under equal per-vector bit budgets. To ensure fairness, we define the bit budget used to represent each vector, which determines its theoretical compression capacity. For PQ, the bit budget equals the number of sub-vectors multiplied by the bits allocated per sub-vector. For RaBitQ, it equals the vector dimensionality multiplied by the bits assigned per dimension (typically one bit for binary quantization). For \ourPQ, the bit budget extends that of PQ by introducing an expansion factor corresponding to the number of refinement epochs, reflecting its progressive quantization mechanism.

\autoref{fig:Quantization_algorithm_comparison} illustrates the trade-off between Recall@10,000 and bit budget across three datasets. Under the same bit budget, \ourPQ consistently achieves higher recall, while requiring one additional FP32 scaling factor per subspace at each refinement epoch. For example, on SIFT1M, at a bit budget of 400 bits per vector, \ourPQ improves recall by 4.1\% and 3.2\% over PQ and RaBitQ, respectively. Similar improvements are observed at lower bit budgets. Even on lower-dimensional datasets such as GloVe, \ourPQ achieves comparable performance to RaBitQ and consistently outperforms PQ.
These results demonstrate \ourPQ's effectiveness across datasets with diverse dimensionalities and data distributions, highlighting the superior quantization performance of \ourPQ due to its direction-aware quantization and progressive refinement.

\subsubsection{Convergence Analysis.} 
\begin{figure}[!t]
    \centering
    \includegraphics[width=\linewidth]
    {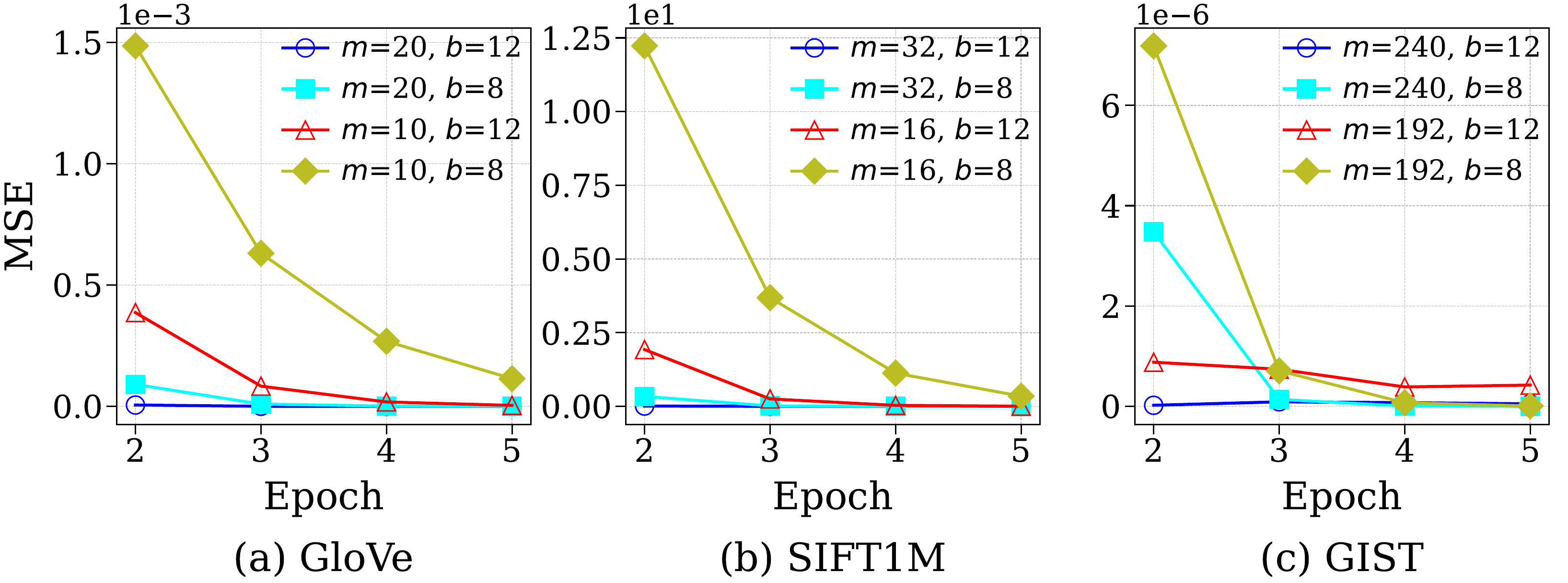}
    \vspace{-0.25in}
    \caption{Convergence analysis of \ourPQ.}
    \vspace{-0.15in}
    \label{fig:DPPQ_mse_epoch}
\end{figure}

We further analyze the effect of hierarchical residual refinement in \ourPQ. \autoref{fig:DPPQ_mse_epoch} reports quantization error (MSE) over refinement epochs under different bit budgets, where $m$ is the number of sub-vectors and $b$ is the number of bits per sub-vector. Across configurations and datasets, the error drops sharply within the first three epochs and stabilizes after only a few iterations, showing that progressive refinement rapidly captures finer-grained directional structure without extensive training. Higher bit budgets further accelerate convergence, and the consistent trends across datasets demonstrate the robustness of the direction-aware progressive optimization strategy.

\subsection{Case Study}
\label{case_study}

To evaluate the real-world impact of \ourSys, we compare with Legacy TEngineDB, StarRocks, and Milvus on \emph{Tencent-Image}, a production-scale dataset deployed in Tencent’s internal image retrieval system. The legacy TEngineDB adopts a traditional segment-coupled indexing framework, where each segment maintains an independent local vector index. As shown in \autoref{fig:case_study}, \ourSys consistently outperforms Legacy TEngineDB, StarRocks, and Milvus for $k=20{,}000$, a typical setting for large-scale exploratory retrieval. At a recall of 0.88, \ourSys achieves a $14\times$, $23\times$, and $4\times$ latency reduction compared to the legacy system, StarRocks, and Milvus, with absolute latencies of 14.2s, 200s, 333s, and 56s, respectively. This performance gap further widens at a 0.8 recall, where \ourSys outperforms the legacy system by $52\times$, StarRocks by $65\times$, and Milvus by $26\times$.
Such an improvement fundamentally shifts large-$k$ retrieval from a latency bottleneck to a practically deployable capability in production. This case study validates that the architectural redesign of \ourSys translates into substantial efficiency gains in real-world deployments.


%% file: Sections/8-Conclusion.tex
\section{Lessons Learned}
\label{sec:lessons}

\textbf{Limitations of Segment-Coupled Indexing.} 
Our experience indicates that segment-coupled indexing is unsuitable for large-$k$ workloads. 
Legacy TEngineDB, built upon this architecture, performed well 
for millions of vectors with hundreds of segments, yet encountered severe scalability bottlenecks 
beyond 1B rows and 2,000 segments. Two fundamental issues drove this limitation: \textit{(1) $k$-amplification overhead} including 
excessive disk I/O, network traffic, and global merge costs. 
\textit{(2) Index cache thrashing.} File-granularity index caching was used to adapt our three-tier storage hierarchy (Remote $\rightarrow$ Local SSD $\rightarrow$ Memory). 
However, this design becomes a liability when SSD and memory capacity are constrained: a single query must access all index files for local searching, triggering frequent index file evictions and reloads that degrade query latency.

\begin{figure}[!t]
    \centering
    \includegraphics[width=0.9\linewidth]{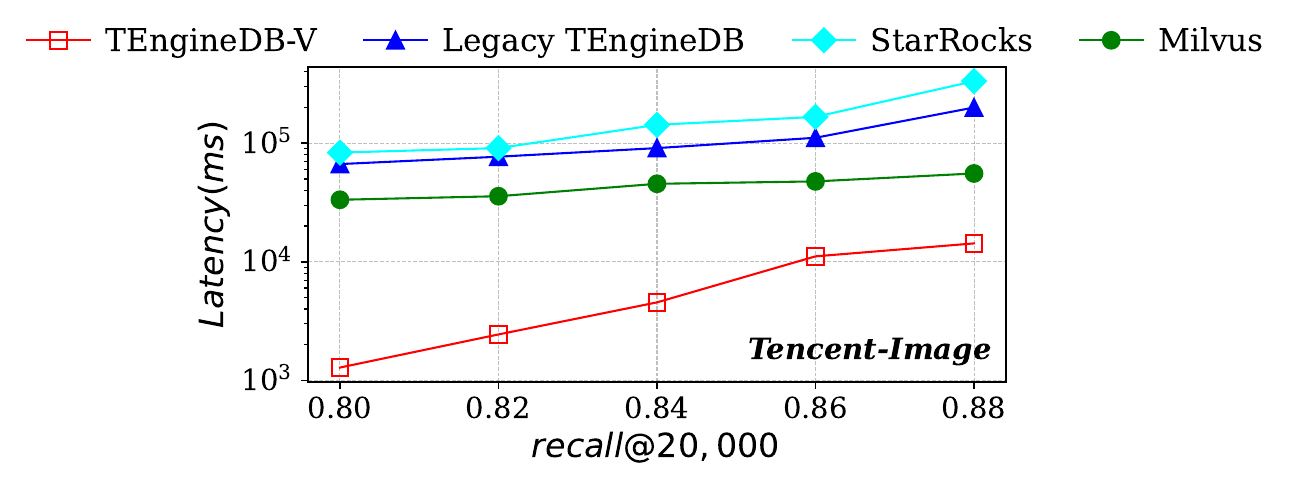}
    \vspace{-0.2in}
    \caption{Production performance at 10-billion scale.}
    \vspace{-0.2in}
    \label{fig:case_study}
\end{figure}

\noindent \textbf{Incremental Optimizations Are Not Enough.}
Three incremental optimizations were investigated to address these issues. First, we increased segment capacity from 2GB to 20GB to reduce $N$ and lower the $N \times k$ amplification. However, this increased write amplification, compaction costs, and data re-distribution overhead during node failures or cluster scaling, rendering it impractical for production. Second, we implemented two-round search: first searching only vector IDs to find the global top-$k$, then fetching payloads. Yet this still requires reading the full vector index from every segment and performing heavy global merges. Third, we tested IVF-cluster-level caching instead of file-level caching, but this only improved intra-segment cache granularity without alleviating thrashing across 2,000+ independent segments. These practical constraints forced us to abandon the segment-coupled paradigm entirely, leading to our segment-decoupled design with table-centric index storage. By maintaining a global index and leveraging the native multi-tier fine-grained cache, we eliminate the $N \times k$ I/O explosion and critical cache thrashing inherent in the coupled design, achieving stable performance regardless of $k$ magnitude.

\noindent \textbf{Complex Similarity Search and Hybrid Workloads.} Our experience deploying \ourSys reveals two trends. First, similarity queries have grown increasingly complex, moving beyond simple top-$k$ retrieval toward complex patterns including similarity-bounded range aggregation~\cite{lan2024cardinality,liang2024unify} and similarity joins \cite{ma2017novel,xie2025fast,chen2025diskjoin}. Second, it is increasingly common to run similarity search and structured analytics simultaneously on the same multi-modal dataset. For instance, our users are performing large-$k$ image retrieval while concurrently generating BI reports on structured image metadata (file size, source, creation time, etc.). By decomposing vector search into relational operators, \ourSys enables unified execution plans in which vector search is jointly optimized with other relational operators such as joins and filters. The optimizer can dynamically reorder, interleave, or push these operators into vector search as runtime filters based on predicate selectivity and execution cost, enabling early pruning and reducing unnecessary candidate materialization. These optimizations extend beyond the FANNS cases in \autoref{tab:costmodel} and generalize to more complex production workloads.

\section{Conclusion}
\label{sec: conclusion}
This paper presents \ourSys, an OLAP-native vector search system tailored for large-$k$ workloads. 
\ourSys establishes a new execution paradigm that materializes segment-decoupled global indexes as relational tables and executes vector search via relational operators. 
Building upon this, we develop \ourPQ to improve quantization accuracy while preserving relational efficiency, design an index-aware distributed optimization framework to ensure efficient execution across diverse workloads, and introduce several OLAP-native optimizations to further accelerate query processing. 
Extensive experiments and real-world production deployments validate the substantial performance gains of \ourSys.
